\documentclass{PoS}

\usepackage{graphicx}
\usepackage{epstopdf}
\usepackage{amsmath,amssymb}
\usepackage{tikz}
\usepackage{etoolbox}

\usetikzlibrary{positioning,arrows,calc,decorations.pathmorphing,decorations.markings}

\newcommand{\beq}{\begin{equation}}
\newcommand{\eeq}{\end{equation}}
\newcommand{\bea}{\begin{eqnarray}}
\newcommand{\eea}{\end{eqnarray}}
\newcommand{\eq}{Eq.~}

\newcommand{\fig}{Fig.~}

\DeclareMathOperator{\tr}{tr}

\def\lsi{\raise0.3ex\hbox{$<$\kern-0.75em\raise-1.1ex\hbox{$\sim$}}}
\def\gsi{\raise0.3ex\hbox{$>$\kern-0.75em\raise-1.1ex\hbox{$\sim$}}}
\newcommand{\losim}{\mathop{\lsi}}

\newenvironment{midtikzpicture}
{\begin{tikzpicture}[baseline={([yshift=-.5ex]current bounding box.center)}]}
{\end{tikzpicture}}

\tikzset{
  circ/.style={%
    black,thick,circle,draw,inner sep=2pt,scale=0.75
  },
  dot/.style={
    black,fill,circle,inner sep=0pt,minimum size=6pt
  },
  inner/.style={
    black,fill,circle,inner sep=0pt,minimum size=3pt
  },
  outer/.style={
    black,draw,thick,circle,inner sep=0pt,minimum size=6pt
  },
  directed/.style={postaction={decorate},
    decoration={markings,mark=at position .65 with{\arrow{stealth}}}},
}

\title{Heavy dense QCD from a 3d effective lattice theory}

\ShortTitle{Effective lattice theory for dense QCD}

\newcommand{\Speaker}

\makeatletter
\renewcommand\speaker[1]{
      \if@speaker%
        \expandafter\def\expandafter\Speaker\expandafter{\Speaker, #1}%
      \else%
        \expandafter\def\expandafter\Speaker\expandafter{#1}%
      \fi%
      \global\setbox\@firstaubox
      \hbox{{\let\thanks\@gobble
        \let\footnote\@gobble\small
        \rm \Speaker}}%
      \if@speaker%
        #1\rlap{\raisebox{-1.7pt}{\textsuperscript{\textnormal{*}}}}%
      \else%
        \global\@speakertrue%
        #1\thanks{Speakers.}%
      \fi
      }%
\makeatother

\author{\speaker{Jonas Glesaaen},\: Mathias Neuman, \speaker{Owe Philipsen}\\
        Institut für Theoretische Physik - Johann Wolfgang Goethe-Universität\\
        Max-von-Laue-Str. 1, 60438 Frankfurt am Main, Germany\\
        E-mail: \email{glesaaen, philipsen@th.physik.uni-frankfurt.de} }

\abstract{%
  The cold and dense regime of the QCD phase diagram is to this day inaccessible to first principle lattice calculations owing to
  the sign problem. Here we present progress of an ongoing effort to probe this particularly difficult regime utilising a
  dimensionally reduced effective lattice theory with a significantly reduced sign problem.  The effective theory is derived by
  combined character and hopping expansion and is valid for heavy quarks near the continuum.  We show an extension of the
  effective theory to order $u^5\kappa^8$ in the cold regime.  A linked cluster expansion is applied to the effective theory
  resulting in a consistent mechanism for handling the effective theory fully ana\-lytically. The new results are consistent with
  the ones from simulations confirming the viability of analytic methods.  Finally we resum the analytical result which doubles
  the convergence region of the expansion.
}

\FullConference{The 33rd International Symposium on Lattice Field Theory\\
  14 -18 July 2015\\
		Kobe International Conference Center, Kobe, Japan}

\begin{document}

\section{Introduction}

The details of the QCD phase diagram are still largely unknown because of the sign problem of Monte Carlo simulations of lattice
QCD. Approximate methods are able to circumvent this problem only for small quark chemical potentials $\mu=\mu_B/3\losim T$
\cite{review}.  So far, no sign of a critical point or a first order phase transition has been found in this controlled region.
Complex Langevin algorithms do not suffer from the sign problem, but in some models converge to incorrect answers or in QCD are
restricted to certain regions in parameter space.  A lot of progress has been made recently, but no true phase transition has been
reported in this approach either \cite{langevin}. 

These difficulties have motivated the development of effective lattice theories, which can either be handled analytically or whose
sign problem is mild enough to simulate the cold and dense region of QCD.  Here we update an ongoing project to describe the cold
and dense regime of QCD by means of a 3d effective theory, which is derived by combined strong coupling and hopping expansions.
The resulting effective theory is valid for the description of very heavy quarks for sufficiently weak couplings, such that a
continuum limit of thermodynamic functions can be attempted.  Its sign problem is mild enough for complex Langevin or even Monte
Carlo simulations to describe cold nuclear matter.  The heavy dense region is also studied by complex Langevin simulations in 4d
\cite{sexty}, for which the effective theory results can serve as a benchmark. Furthermore, the effective theory approach can be
tested against full simulations in two-colour QCD \cite{scior,scior2}.  In a new development, we show how the thermodynamic
functions of the effective theory can also be computed fully analytically by means of linked cluster expansion. 

\section{Derivation of the effective theory}

We start on a $(3+1)$-dimensional lattice with Wilson's gauge and fermion actions on infinite spatial volume and  at finite
temperature $T=(aN_\tau)^{-1}$. After Grassmann integration the partition function is
\begin{eqnarray}
  Z=\int[dU_\mu]\det\left[Q\right]^{N_f}\exp\left[S_g\right]\;,\qquad
  S_g=\frac{\beta}{2N_c}\sum_p\left[\tr\,   U_p+\tr\,   U_p^\dagger\right]\;,
\end{eqnarray}
with the quark hopping matrix 
\begin{eqnarray}
  Q^{ab}_{\alpha\beta,xy}=\delta^{ab}\delta_{\alpha\beta}\delta_{xy}-
  \kappa\sum_{\nu=0}^3\left[e^{a\mu\delta_{\nu0}}(1+\gamma_\nu)_{\alpha\beta}U_\nu^{ab}(x)
    \delta_{x,y-\hat{\nu}}+e^{-a\mu\delta_{\nu0}}(1-\gamma_\nu)_{\alpha\beta}U_{-\nu}^{ab}(x)
    \delta_{x,y+\hat{\nu}}\right]
  \;.\nonumber
\end{eqnarray}
The lattice gauge coupling and hopping parameter are defined as
\begin{equation}
  \beta=\frac{2N_C}{g^2},\quad \kappa=\frac{1}{2am+8}\;.
\end{equation}
A 3d effective action is then defined by integrating out the spatial link variables
\begin{eqnarray}
  Z&=&\int[dU_0]e^{-S_\text{eff}[U_0]}
  =\int[dW]e^{-S_\text{eff}[W]},\quad
  e^{S_{\mathrm{eff}}}\equiv\int[dU_k]\det\left[Q\right]^{N_f}\exp\left[S_g\right]\;.
  \label{eq_defeffth}
\end{eqnarray}
The resulting effective theory depends on temporal link variables closing through the temporal boundary, i.e.~on Polyakov loops
\begin{eqnarray}
  W_i\equiv\prod_{\tau=1}^{N_\tau}U_0\left(\vec{x}_i;\tau\right),\quad L_i = \tr W_i\:.
\end{eqnarray}
Without truncations, the effective action is unique and exact.  Since all spatial links are integrated over, the resulting
effective action has long-range interactions of Polyakov loops at all distances and to all powers so that in practice truncations
are necessary.  For non-perturbative ways to determine truncated theories, see \cite{wozar,green1,green2,bergner}.  Here we expand
eq.~(\ref{eq_defeffth}) in a combined strong coupling and hopping expansion, with interaction terms ordered according to their
leading powers in $\beta, \kappa$. 

For the gauge action it is advantageous to perform a character expansion
\begin{eqnarray}
  \exp\left[\frac{\beta}{2N_c}\Big(\tr\,   U+\tr\,   U^\dagger\Big)\right]
    =c_0(\beta)\left[1+\sum_{r\neq0}d_ra_r(\beta)\chi_r(U)\right]\;,
\end{eqnarray}
where the factor $c_0(\beta)$ can be neglected as it is independent of gauge links and cancels in expectation values. In earlier
publications \cite{Langelage:2010yr,Langelage:2010nj}, we have shown how to compute the effective gauge theory up to rather high
orders in the fundamental character expansion coefficient $u(\beta)\equiv a_f(\beta)$. Note that $u(\beta)=\beta/18+O(\beta^2)<1$,
thus improving the convergence compared to straightforward expansion in $\beta$.  To leading order the effective gauge action
reads
\begin{eqnarray}
  e^{S_{\mathrm{eff}}^{(1)}}=\lambda(u,N_\tau)\sum_{<ij>}\left(L_iL_j^\ast+L_i^\ast L_j\right)\;,
  \qquad\lambda(u,N_\tau)=u^{N_\tau}\Big[1+\ldots\Big]\;,
  \label{eq_seffgauge}
\end{eqnarray}
where higher order corrections of $\lambda(u,N_\tau)$ as well as a discussion of higher order interaction terms can be found in
\cite{Langelage:2010yr}. Going via an effective action results in a resummation with better convergence properties than a direct
series expansion of the thermodynamic observables as in \cite{lange1,lange2}.  We observe that $\lambda(u,N_\tau)$ is suppressed
for large $N_\tau$. In this presentation our main interest is in the cold and dense region, for which all $\lambda_i<10^{ -18}$
and the pure gauge contribution can be safely neglected.

To compute the quark determinant in a spatial hopping expansion, we split the quark matrix in positive and negative temporal and
spatial parts,
\begin{eqnarray}
  Q=1-T-S=1-T^+-T^--S^+-S^-\;,
\end{eqnarray}
The static determinant is given by neglecting the spatial parts and can be computed exactly \cite{hetrick}
\begin{eqnarray}
  \det[Q_{\mathrm{stat}}]&=& \prod_{\vec{x}} \det \Big[1+h_1W({\vec{x}})\Big]^2
    \det \Big[1+\bar{h}_1W^{\dagger}({\vec{x}})\Big]^2 \nonumber\\
  &=&\prod_{\vec{x}} \left[1 + h_1 L_{\vec{x}} + h_1^2 L^{\dagger}_{\vec{x}}+h_1^3\right]^2
    \left[1 + \bar{h}_1 L^{\dagger}_{\vec{x}} + \bar{h}_1^2 L_{\vec{x}}+\bar{h}_1^3\right]^2,
  \label{q_static}
\end{eqnarray} 
with the leading order couplings
\begin{equation}
  h_1=(2 \kappa e^{a \mu})^{N_{\tau}},\quad \bar{h}_1=(2 \kappa e^{-a \mu})^{N_{\tau}}\;.
\end{equation}
In order to compute a systematic hopping expansion, we define the kinetic quark determinant
\begin{eqnarray}
  \det[Q]&\equiv&\det[Q_{\mathrm{stat}}]\det[Q_{\mathrm{kin}}]\;,\\
  \det[Q_{\mathrm{kin}}]&=&[1-(1-T)^{-1}(S^++S^-)]\equiv\det[1-P-M]=\exp\left[\tr\,  \ln (1-P-M)\right]\;.
  \label{eq_detqkin}
\end{eqnarray}
The static propagator $(1-T)^{-1}$ is also known exactly.  It contains summation of all temporal windings to produce the basic
building blocks of the effective action, for details see \cite{ka4},
\begin{equation}
  W_{n,m}(\vec{x}) = \tr \, \frac{\big(h_1 W(\vec{x})\big)^m}{\big(1+h_1 W(\vec{x})\big)^n} \, .
\end{equation}
We have calculated the effective action through order $\kappa^8 u^5$ in the low temperature limit, i.e.~the leading power of
$N_\tau$. However, because of its length we will give the result only in a compact, graphical representation.  We symbolise
factors of $W_{n,m}(\vec{x})$ by vertices, where $n$ is the number of bonds entering a vertex, and $m$ is the number indicated on
the node. Furthermore, vertices which are connected by one or more bonds are nearest neighbours on the lattice. 
{ \allowdisplaybreaks
\begin{align} \label{eq:effective_action}
  S_{\mathrm{eff}} &=
  h_2 N_f \sum_{\mathrm{dof}} \; \begin{midtikzpicture}
    \node[circ] (n1) {1};
    \node[circ] (n2) at ([shift={(270:.75)}] n1) {1}
      edge[thick] (n1);
  \end{midtikzpicture}
  \; - h_2^2 N_f\sum_{\mathrm{dof}}  \: \begin{midtikzpicture}
    \node[circ] (n1) {1};
    \node[circ] (n2) at ([shift={(60:{sqrt(2/3)})}] n1) {1}
      edge[thick] (n1);
    \node[circ] (n3) at ([shift={(300:{sqrt(2/3)})}] n2) {1}
      edge[thick] (n2);
  \end{midtikzpicture}
  - h_2^2 N_f^2\sum_{\mathrm{dof}} \: \begin{midtikzpicture}
    \node[circ] (n1) {1};
    \node[circ] (n2) at ([shift={(270:.75)}] n1) {1}
      edge[thick,bend left=45] (n1)
      edge[thick,bend right=45] (n1);
  \end{midtikzpicture}
  + h_2^3 N_f\sum_{\mathrm{dof}} \: \begin{midtikzpicture}
    \node[circ] (n1) {1};
    \node[circ] (n2) at ([shift={(60:{sqrt(2/3)})}] n1) {1}
      edge[thick] (n1);
    \node[circ] (n3) at ([shift={(300:{sqrt(2/3)})}] n2) {1}
      edge[thick] (n2);
    \node[circ] (n4) at ([shift={(60:{sqrt(2/3)})}] n3) {1}
      edge[thick] (n3);
  \end{midtikzpicture} \nonumber\\
  &+\frac{1}{3} h_2^3 N_f \sum_{\mathrm{dof}} \Bigg(\; \begin{midtikzpicture}
    \node[circ] (n1) {1};
    \node[circ] at ([shift={(-30:.5)}] n1) {1}
      edge[thick] (n1);
    \node[circ] at ([shift={(90:.5)}] n1) {1}
      edge[thick] (n1);
    \node[circ] at ([shift={(210:.5)}] n1) {1}
      edge[thick] (n1);
  \end{midtikzpicture} 
  \;-\;\begin{midtikzpicture}
    \node[circ] (n1) {2};
    \node[circ] at ([shift={(-30:.5)}] n1) {1}
      edge[thick] (n1);
    \node[circ] at ([shift={(90:.5)}] n1) {1}
      edge[thick] (n1);
    \node[circ] at ([shift={(210:.5)}] n1) {1}
      edge[thick] (n1);
  \end{midtikzpicture} \;\Bigg)
  +2 h_2^3 N_f^2 \sum_{\mathrm{dof}} \Bigg(\; \begin{midtikzpicture}
    \node[circ] (n1) {1};
    \node[circ] (n2) at ([shift={(60:.75)}] n1) {1}
      edge[thick,bend left=30] (n1)
      edge[thick,bend right=30] (n1);
    \node[circ] at ([shift={(300:.75)}] n2) {1}
      edge[thick] (n2);
  \end{midtikzpicture} 
  \;-\; \begin{midtikzpicture}
    \node[circ] (n1) {1};
    \node[circ] (n2) at ([shift={(60:.75)}] n1) {2}
      edge[thick,bend left=30] (n1)
      edge[thick,bend right=30] (n1);
    \node[circ] at ([shift={(300:.75)}] n2) {1}
      edge[thick] (n2);
  \end{midtikzpicture} \Bigg) \nonumber\\
  &+\frac{1}{6} h_2^3 N_f \sum_{\mathrm{dof}} \Bigg(\; \begin{midtikzpicture}
    \node[circ] (n1) {1};
    \node[circ] at ([shift={(270:.75)}] n1) {1}
      edge[thick,bend left=45] (n1)
      edge[thick,bend right=45] (n1)
      edge[thick] (n1);
  \end{midtikzpicture} 
  \;-\; \begin{midtikzpicture}
    \node[circ] (n1) {2};
    \node[circ] at ([shift={(270:.75)}] n1) {2}
      edge[thick,bend left=45] (n1)
      edge[thick,bend right=45] (n1)
      edge[thick] (n1);
  \end{midtikzpicture} \Bigg)
  - \frac{4}{3} h_2^3 N_f^3 \sum_{\mathrm{dof}} \; \begin{midtikzpicture}
    \node[circ] (n1) {1};
    \node[circ] at ([shift={(270:.75)}] n1) {2}
      edge[thick,bend left=45] (n1)
      edge[thick,bend right=45] (n1)
      edge[thick] (n1);
  \end{midtikzpicture} 
  - h_2^4 N_f\sum_{\mathrm{dof}} \: \begin{midtikzpicture}
    \node[circ] (n1) {1};
    \node[circ] (n2) at ([shift={(60:{sqrt(2/3)})}] n1) {1}
      edge[thick] (n1);
    \node[circ] (n3) at ([shift={(300:{sqrt(2/3)})}] n2) {1}
      edge[thick] (n2);
    \node[circ] (n4) at ([shift={(60:{sqrt(2/3)})}] n3) {1}
      edge[thick] (n3);
    \node[circ] (n5) at ([shift={(300:{sqrt(2/3)})}] n4) {1}
      edge[thick] (n4);
  \end{midtikzpicture} \nonumber\\
  &- \frac{1}{12} h_2^4 N_f\sum_{\mathrm{dof}} \Bigg(\; \begin{midtikzpicture}
    \node[circ] (n1) {1};
    \node[circ] at ([shift={(45:.5)}] n1) {1}
      edge[thick] (n1);
    \node[circ] at ([shift={(135:.5)}] n1) {1}
      edge[thick] (n1);
    \node[circ] at ([shift={(225:.5)}] n1) {1}
      edge[thick] (n1);
    \node[circ] at ([shift={(315:.5)}] n1) {1}
      edge[thick] (n1);
  \end{midtikzpicture}
  \;-2\; \begin{midtikzpicture}
    \node[circ] (n1) {2};
    \node[circ] at ([shift={(45:.5)}] n1) {1}
      edge[thick] (n1);
    \node[circ] at ([shift={(135:.5)}] n1) {1}
      edge[thick] (n1);
    \node[circ] at ([shift={(225:.5)}] n1) {1}
      edge[thick] (n1);
    \node[circ] at ([shift={(315:.5)}] n1) {1}
      edge[thick] (n1);
  \end{midtikzpicture}
  \;+\; \begin{midtikzpicture}
    \node[circ] (n1) {3};
    \node[circ] at ([shift={(45:.5)}] n1) {1}
      edge[thick] (n1);
    \node[circ] at ([shift={(135:.5)}] n1) {1}
      edge[thick] (n1);
    \node[circ] at ([shift={(225:.5)}] n1) {1}
      edge[thick] (n1);
    \node[circ] at ([shift={(315:.5)}] n1) {1}
      edge[thick] (n1);
  \end{midtikzpicture} \Bigg)
  -  h_2^4 N_f\sum_{\mathrm{dof}} \Bigg(\; \begin{midtikzpicture}
    \node[circ] (n1) {1};
    \node[circ] (n2) at ([shift={(60:{sqrt(1/3)})}] n1) {1}
      edge[thick] (n1);
    \node[circ] (n3) at ([shift={(300:{sqrt(1/3)})}] n2) {1}
      edge[thick] (n2);
    \node[circ] (n4) at ([shift={(60:{sqrt(1/3)})}] n3) {1}
      edge[thick] (n3);
    \node[circ] (n5) at ([shift={(0:.5)}] n3) {1}
      edge[thick] (n3);
  \end{midtikzpicture}
  \;-\; \begin{midtikzpicture}
    \node[circ] (n1) {1};
    \node[circ] (n2) at ([shift={(60:{sqrt(1/3)})}] n1) {1}
      edge[thick] (n1);
    \node[circ] (n3) at ([shift={(300:{sqrt(1/3)})}] n2) {2}
      edge[thick] (n2);
    \node[circ] (n4) at ([shift={(60:{sqrt(1/3)})}] n3) {1}
      edge[thick] (n3);
    \node[circ] (n5) at ([shift={(0:.5)}] n3) {1}
      edge[thick] (n3);
  \end{midtikzpicture} \Bigg) \nonumber\\
  &- h_2^4 N_f^2 \sum_{\mathrm{dof}} \Bigg(\; \begin{midtikzpicture}
    \node[circ] (n1) {1};
    \node[circ] at ([shift={(-30:.5)}] n1) {1}
      edge[thick] (n1);
    \node[circ] at ([shift={(90:.65)}] n1) {1}
      edge[thick,bend left=30] (n1)
      edge[thick,bend right=30] (n1);
    \node[circ] at ([shift={(210:.5)}] n1) {1}
      edge[thick] (n1);
  \end{midtikzpicture} 
  -4\begin{midtikzpicture}
    \node[circ] (n1) {2};
    \node[circ] at ([shift={(-30:.5)}] n1) {1}
      edge[thick] (n1);
    \node[circ] at ([shift={(90:.65)}] n1) {1}
      edge[thick,bend left=30] (n1)
      edge[thick,bend right=30] (n1);
    \node[circ] at ([shift={(210:.5)}] n1) {1}
      edge[thick] (n1);
  \end{midtikzpicture}
  +\begin{midtikzpicture}
    \node[circ] (n1) {3};
    \node[circ] at ([shift={(-30:.5)}] n1) {1}
      edge[thick] (n1);
    \node[circ] at ([shift={(90:.65)}] n1) {1}
      edge[thick,bend left=30] (n1)
      edge[thick,bend right=30] (n1);
    \node[circ] at ([shift={(210:.5)}] n1) {1}
      edge[thick] (n1);
  \end{midtikzpicture} \;\Bigg)
  - h_2^4 N_f^2 \sum_{\mathrm{dof}} \begin{midtikzpicture}
    \node[circ] (n1) {1};
    \node[circ] (n2) at (0.65,0) {1}
      edge[thick] (n1);
    \node[circ] (n3) at (0.65,0.65) {1}
      edge[thick] (n2);
    \node[circ] (n4) at (0,0.65) {1}
      edge[thick] (n3)
      edge[thick] (n1);
  \end{midtikzpicture} \nonumber\\
  &-2 h_2^4 N_f^2 \sum_{\mathrm{dof}} \Bigg( \; \begin{midtikzpicture}
    \node[circ] (n1) {1};
    \node[circ] (n2) at ([shift={(60:.75)}] n1) {1}
      edge[thick,bend left=30] (n1)
      edge[thick,bend right=30] (n1);
    \node[circ] (n3) at ([shift={(300:.75)}] n2) {1}
      edge[thick] (n2);
    \node[circ] at ([shift={(60:.75)}] n3) {1}
      edge[thick] (n3);
  \end{midtikzpicture}
  \;-\; \begin{midtikzpicture}
    \node[circ] (n1) {1};
    \node[circ] (n2) at ([shift={(60:.75)}] n1) {2}
      edge[thick,bend left=30] (n1)
      edge[thick,bend right=30] (n1);
    \node[circ] (n3) at ([shift={(300:.75)}] n2) {1}
      edge[thick] (n2);
    \node[circ] at ([shift={(60:.75)}] n3) {1}
      edge[thick] (n3);
  \end{midtikzpicture} \Bigg)
  - h_2^4 N_f^2 \sum_{\mathrm{dof}} \Bigg( \; \begin{midtikzpicture}
    \node[circ] (n1) {1};
    \node[circ] (n2) at ([shift={(65:{sqrt(2/3)})}] n1) {1}
      edge[thick] (n1);
    \node[circ] (n3) at ([shift={(-65:{sqrt(2/3)})}] n2) {1}
      edge[thick,bend left=30] (n2)
      edge[thick,bend right=30] (n2);
    \node[circ] (n4) at ([shift={(65:{sqrt(2/3)})}] n3) {1}
      edge[thick] (n3);
  \end{midtikzpicture}
  \;-2\; \begin{midtikzpicture}
    \node[circ] (n1) {1};
    \node[circ] (n2) at ([shift={(65:{sqrt(2/3)})}] n1) {1}
      edge[thick] (n1);
    \node[circ] (n3) at ([shift={(-65:{sqrt(2/3)})}] n2) {2}
      edge[thick,bend left=30] (n2)
      edge[thick,bend right=30] (n2);
    \node[circ] (n4) at ([shift={(65:{sqrt(2/3)})}] n3) {1}
      edge[thick] (n3);
  \end{midtikzpicture}
  \;+\; \begin{midtikzpicture}
    \node[circ] (n1) {1};
    \node[circ] (n2) at ([shift={(65:{sqrt(2/3)})}] n1) {2}
      edge[thick] (n1);
    \node[circ] (n3) at ([shift={(-65:{sqrt(2/3)})}] n2) {2}
      edge[thick,bend left=30] (n2)
      edge[thick,bend right=30] (n2);
    \node[circ] (n4) at ([shift={(65:{sqrt(2/3)})}] n3) {1}
      edge[thick] (n3);
  \end{midtikzpicture} \Bigg) \nonumber\\
  &-\frac{1}{3} h_2^4 N_f \sum_{\mathrm{dof}} \Bigg( \; \begin{midtikzpicture}
    \node[circ] (n1) {1};
    \node[circ] (n2) at ([shift={(60:.75)}] n1) {1}
      edge[thick,bend left=30] (n1)
      edge[thick,bend right=30] (n1)
      edge[thick] (n1);
    \node[circ] (n3) at ([shift={(300:.75)}] n2) {1}
      edge[thick] (n2);
  \end{midtikzpicture}
  \;-2\; \begin{midtikzpicture}
    \node[circ] (n1) {1};
    \node[circ] (n2) at ([shift={(60:.75)}] n1) {2}
      edge[thick,bend left=30] (n1)
      edge[thick,bend right=30] (n1)
      edge[thick] (n1);
    \node[circ] (n3) at ([shift={(300:.75)}] n2) {1}
      edge[thick] (n2);
  \end{midtikzpicture}
  \;+2\; \begin{midtikzpicture}
    \node[circ] (n1) {2};
    \node[circ] (n2) at ([shift={(60:.75)}] n1) {2}
      edge[thick,bend left=30] (n1)
      edge[thick,bend right=30] (n1)
      edge[thick] (n1);
    \node[circ] (n3) at ([shift={(300:.75)}] n2) {1}
      edge[thick] (n2);
  \end{midtikzpicture}
  \;-\; \begin{midtikzpicture}
    \node[circ] (n1) {2};
    \node[circ] (n2) at ([shift={(60:.75)}] n1) {3}
      edge[thick,bend left=30] (n1)
      edge[thick,bend right=30] (n1)
      edge[thick] (n1);
    \node[circ] (n3) at ([shift={(300:.75)}] n2) {1}
      edge[thick] (n2);
  \end{midtikzpicture} \Bigg) \nonumber\\
  &+\frac{4}{3} h_2^4 N_f^3 \sum_{\mathrm{dof}} \Bigg( \; \begin{midtikzpicture}
    \node[circ] (n1) {2};
    \node[circ] (n2) at ([shift={(60:.75)}] n1) {1}
      edge[thick,bend left=30] (n1)
      edge[thick,bend right=30] (n1)
      edge[thick] (n1);
    \node[circ] (n3) at ([shift={(300:.75)}] n2) {1}
      edge[thick] (n2);
  \end{midtikzpicture}
  \;-2\; \begin{midtikzpicture}
    \node[circ] (n1) {2};
    \node[circ] (n2) at ([shift={(60:.75)}] n1) {2}
      edge[thick,bend left=30] (n1)
      edge[thick,bend right=30] (n1)
      edge[thick] (n1);
    \node[circ] (n3) at ([shift={(300:.75)}] n2) {1}
      edge[thick] (n2);
  \end{midtikzpicture}
  \;+2\; \begin{midtikzpicture}
    \node[circ] (n1) {1};
    \node[circ] (n2) at ([shift={(60:.75)}] n1) {2}
      edge[thick,bend left=30] (n1)
      edge[thick,bend right=30] (n1)
      edge[thick] (n1);
    \node[circ] (n3) at ([shift={(300:.75)}] n2) {1}
      edge[thick] (n2);
  \end{midtikzpicture}
  \;-\; \begin{midtikzpicture}
    \node[circ] (n1) {1};
    \node[circ] (n2) at ([shift={(60:.75)}] n1) {3}
      edge[thick,bend left=30] (n1)
      edge[thick,bend right=30] (n1)
      edge[thick] (n1);
    \node[circ] (n3) at ([shift={(300:.75)}] n2) {1}
      edge[thick] (n2);
  \end{midtikzpicture} \Bigg) \nonumber\\
  &-\bigg(\frac{1}{12} N_f + \frac{2}{3} N_f^3 \bigg) h_2^4 \sum_{\mathrm{dof}} \Bigg( \; \begin{midtikzpicture}
    \node[circ] (n1) {1};
    \node[circ] (n2) at ([shift={(60:.75)}] n1) {1}
      edge[thick,bend left=30] (n1)
      edge[thick,bend right=30] (n1);
    \node[circ] (n3) at ([shift={(300:.75)}] n2) {1}
      edge[thick,bend left=30] (n2)
      edge[thick,bend right=30] (n2);
  \end{midtikzpicture}
  \;-4\; \begin{midtikzpicture}
    \node[circ] (n1) {1};
    \node[circ] (n2) at ([shift={(60:.75)}] n1) {2}
      edge[thick,bend left=30] (n1)
      edge[thick,bend right=30] (n1);
    \node[circ] (n3) at ([shift={(300:.75)}] n2) {1}
      edge[thick,bend left=30] (n2)
      edge[thick,bend right=30] (n2);
  \end{midtikzpicture}
  \;+\; \begin{midtikzpicture}
    \node[circ] (n1) {1};
    \node[circ] (n2) at ([shift={(60:.75)}] n1) {3}
      edge[thick,bend left=30] (n1)
      edge[thick,bend right=30] (n1);
    \node[circ] (n3) at ([shift={(300:.75)}] n2) {1}
      edge[thick,bend left=30] (n2)
      edge[thick,bend right=30] (n2);
  \end{midtikzpicture} \Bigg)
  -\frac{2}{3} h_2^4 N_f^4 \sum_{\mathrm{dof}} \Bigg( \; \begin{midtikzpicture}
    \node[circ] (n1) {1};
    \node[circ] (n2) at (0,-.75) {3}
      edge[thick, bend right=45] (n1)
      edge[thick, bend left=45] (n1)
      edge[thick, bend right=15] (n1)
      edge[thick, bend left=15] (n1);
  \end{midtikzpicture}
  \;+2\; \begin{midtikzpicture}
    \node[circ] (n1) {2};
    \node[circ] (n2) at (0,-.75) {2}
      edge[thick, bend right=45] (n1)
      edge[thick, bend left=45] (n1)
      edge[thick, bend right=15] (n1)
      edge[thick, bend left=15] (n1);
  \end{midtikzpicture} \Bigg) \nonumber \\
  &-\frac{1}{12} h_2^4 N_f^2 \sum_{\mathrm{dof}} \Bigg( \; \begin{midtikzpicture}
    \node[circ] (n1) {1};
    \node[circ] (n2) at (0,-.75) {1}
      edge[thick, bend right=45] (n1)
      edge[thick, bend left=45] (n1)
      edge[thick, bend right=15] (n1)
      edge[thick, bend left=15] (n1);
  \end{midtikzpicture}
  \;+12\; \begin{midtikzpicture}
    \node[circ] (n1) {2};
    \node[circ] (n2) at (0,-.75) {2}
      edge[thick, bend right=45] (n1)
      edge[thick, bend left=45] (n1)
      edge[thick, bend right=15] (n1)
      edge[thick, bend left=15] (n1);
  \end{midtikzpicture}
  \;+\; \begin{midtikzpicture}
    \node[circ] (n1) {3};
    \node[circ] (n2) at (0,-.75) {3}
      edge[thick, bend right=45] (n1)
      edge[thick, bend left=45] (n1)
      edge[thick, bend right=15] (n1)
      edge[thick, bend left=15] (n1);
  \end{midtikzpicture} \Bigg)
  +\frac{2}{3} h_2^4 N_f^2 \sum_{\mathrm{dof}} \Bigg( \; \begin{midtikzpicture}
    \node[circ] (n1) {1};
    \node[circ] (n2) at (0,-.75) {2}
      edge[thick, bend right=45] (n1)
      edge[thick, bend left=45] (n1)
      edge[thick, bend right=15] (n1)
      edge[thick, bend left=15] (n1);
  \end{midtikzpicture}
  \;+\; \begin{midtikzpicture}
    \node[circ] (n1) {2};
    \node[circ] (n2) at (0,-.75) {3}
      edge[thick, bend right=45] (n1)
      edge[thick, bend left=45] (n1)
      edge[thick, bend right=15] (n1)
      edge[thick, bend left=15] (n1);
  \end{midtikzpicture} \Bigg) +\mathcal{O} \big(\kappa^{10}\big)
\end{align}
}
The effective couplings to this order are
\begin{eqnarray}
  h_1&=& e^{N_{\tau} (a \mu + \log(2 \kappa))} e^{6 N_{\tau} \kappa^2 u (\frac{1}{1-u}
  +4u^4-12\kappa^2+9\kappa^2 u +4\kappa^2 u^2-4\kappa^4)}, \\
  h_2&=& \frac{\kappa^2 N_{\tau}}{N_c} \Big[1 + \frac{2 u}{1-u} + 8 u^5 \Big]   \;.
\end{eqnarray}
The sign problem of this effective theory is weak and it can be simulated either by complex Langevin without any convergence or
runaway problems, or even by standard Metropolis using reweighting. For details and tests of the simulation algorithm see
\cite{ka4}.

\section{The QCD phase diagram for heavy quarks}

\begin{figure}[t]
  \vspace*{-0.3cm}
  \includegraphics[width=0.5\textwidth]{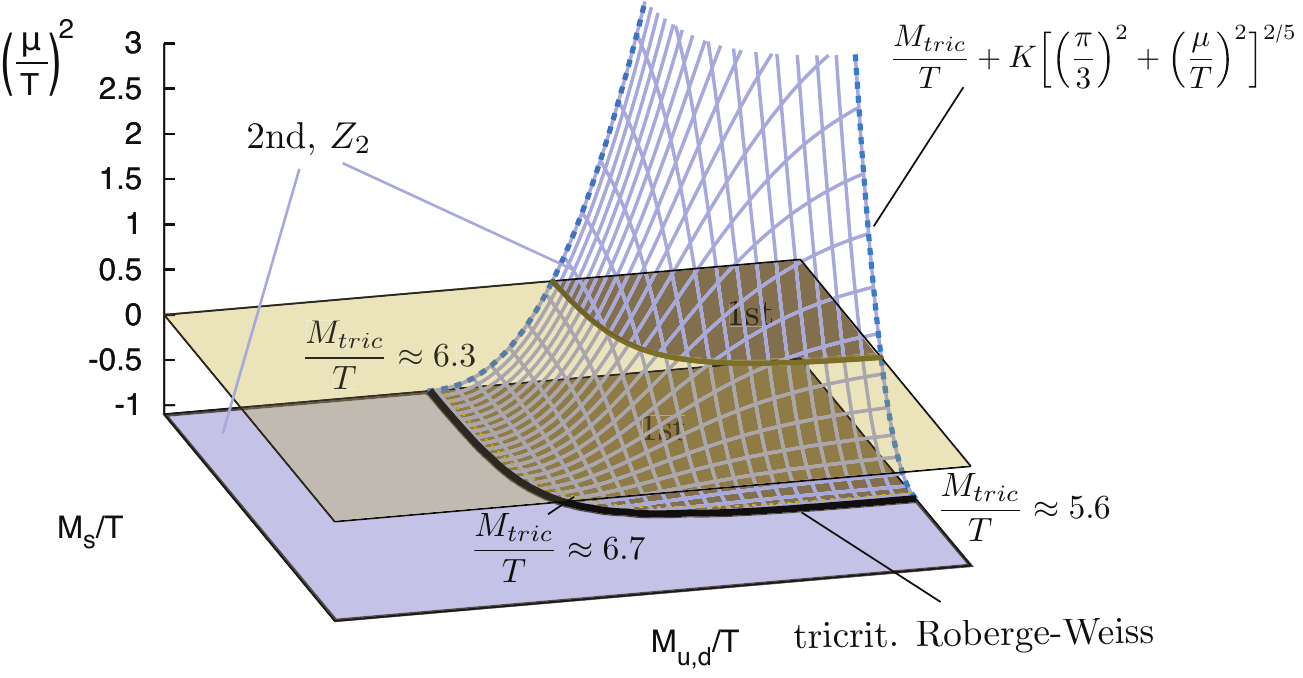}
  \includegraphics[width=0.4\textwidth]{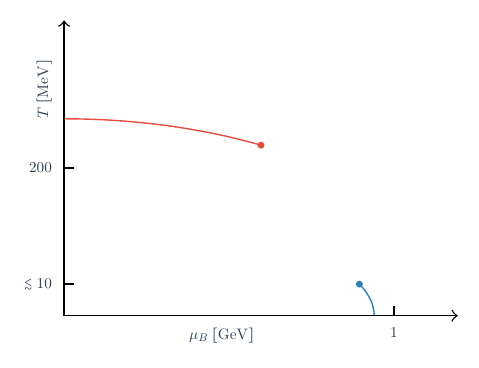}
  \caption[]{Left: The deconfinement critical surface with heavy quarks for real and imaginary chemical 
    potential. From \cite{Fromm:2011qi}. Right: The qualitative phase diagram for QCD with heavy quarks. The location of the lines
    and critical endpoints depends on $N_f$ and the quark mass.}
  \label{fig:finiteT}
\end{figure}
The effective theory has been used to study the QCD phase diagram for QCD with heavy quarks. In the static limit, the finite
temperature deconfinement transition is of first order, corresponding to the spontaneous breaking of centre symmetry.  With
dynamical quarks, the fermion determinant breaks the symmetry explicitly  and the transition weakens until it changes to a
crossover at a critical quark mass. Similarly, real chemical potential weakens the transition. This behaviour is reflected in the
heavy mass corner of the Columbia plot.  \fig\ref{fig:finiteT} (left) shows the deconfinement critical surface separating the
first order from the crossover region as calculated with a $\kappa^2$-action but full $N_\tau$- dependence in \cite{Fromm:2011qi}. 

\begin{figure}[t]
  \vspace*{-0.3cm}
  \includegraphics[width=0.33\textwidth]{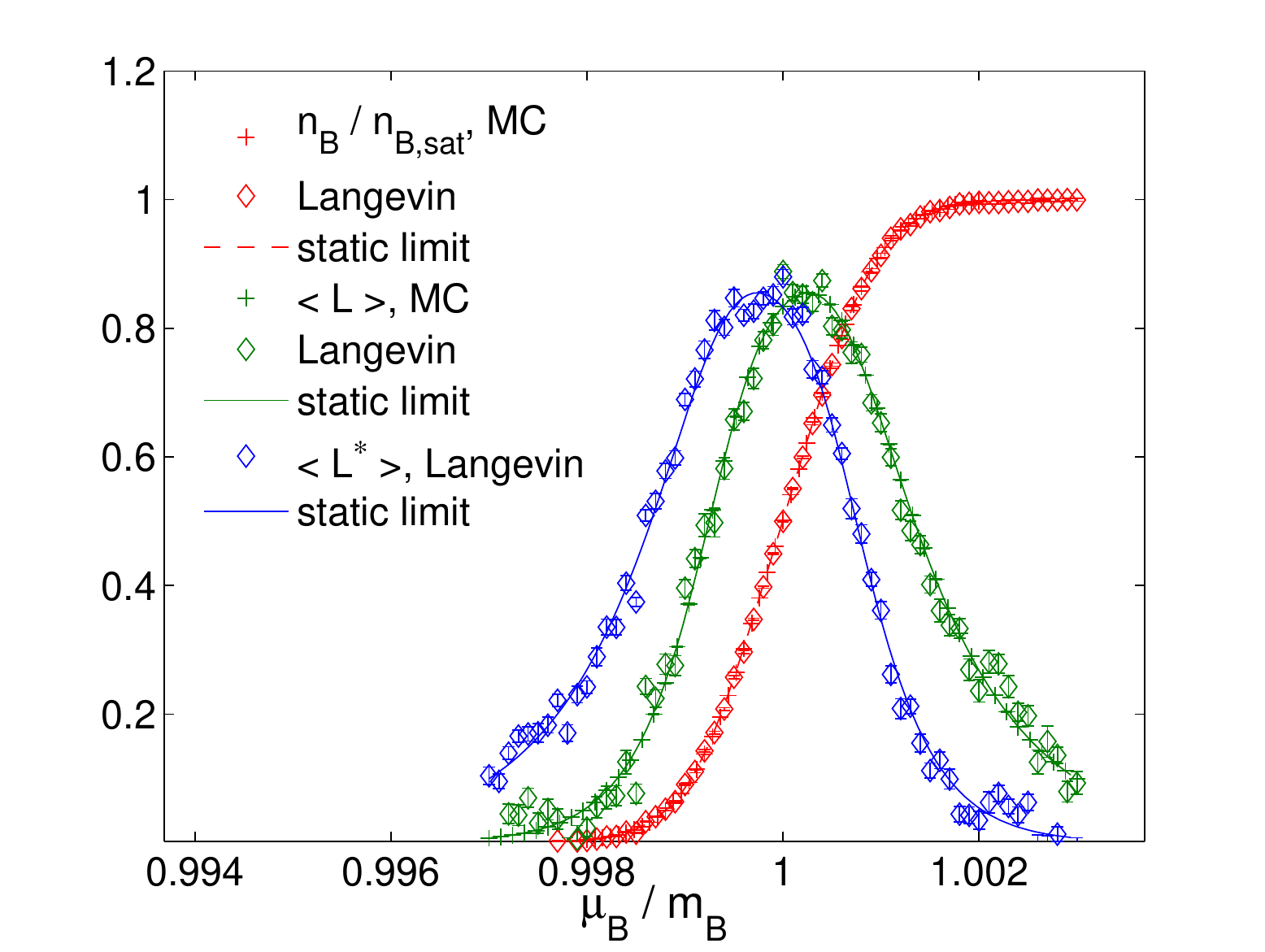}
  \includegraphics[width=0.33\textwidth]{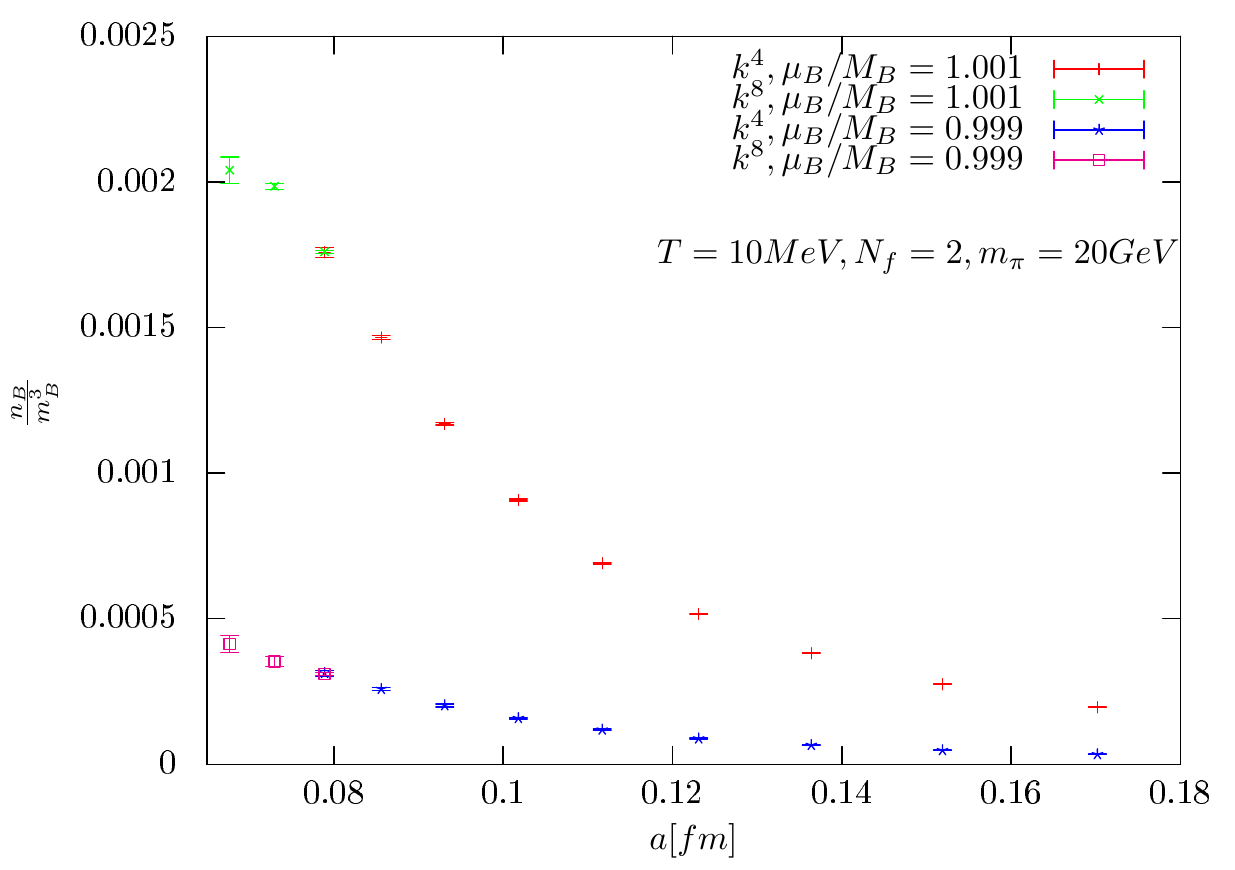}
  \includegraphics[width=0.33\textwidth]{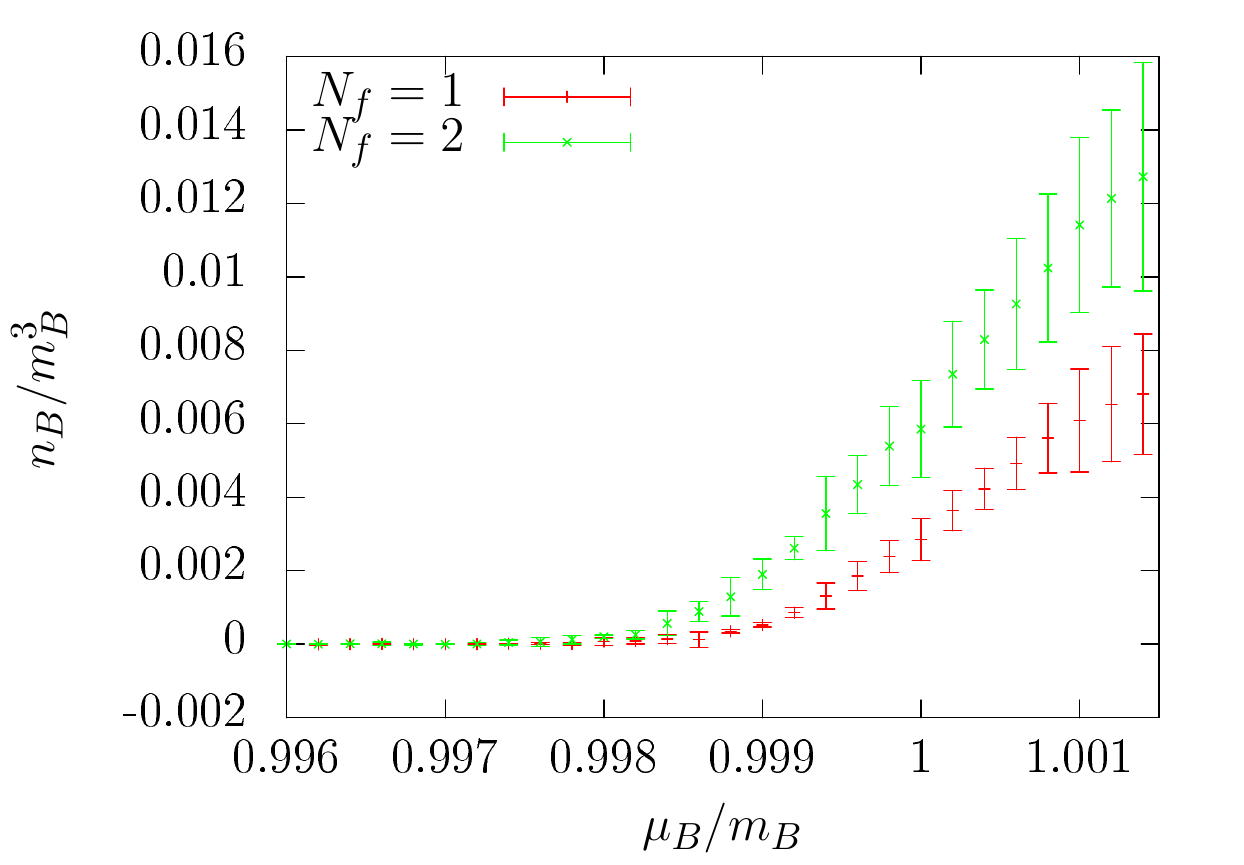}
  \caption[]{Left: Baryon density, Polyakov loop and conjugate Polyakov loop 
    obtained from Monte Carlo $N_s=3$, complex Langevin ($N_s=6$) and the static strong coupling limit, respectively. From
    \cite{Fromm:2012eb}. Middle: Continuum extrapolation for two densities.  Right: Baryon density in the continuum. From
    \cite{ka4}.}
  \label{fig:dense}
\end{figure}

The cold and dense regime was considered in \cite{Fromm:2012eb,ka4}. \fig\ref{fig:dense} (left) shows results at a fixed lattice
spacing.  We set the scale using the $r_0$ parameter and compute the corresponding pion mass from strong coupling formulae
\cite{ka4}.  To keep our truncated series in full control, we choose $\beta=5.7, \kappa = 0.0000887, N_\tau=116$ corresponding to
$m_\pi=20$ GeV, $T=10$ MeV, $a=0.17$ fm.  The silver blaze property as well as lattice saturation are clearly observed.  Note that
the Polyakov loop as well as its conjugate get screened in the presence of a baryonic medium, and hence rise.  The ensuing
decrease is due to the artefact of lattice saturation which forces all $Z(3)$ states to be occupied. Note that this decrease
happens before saturation near the point of half filling. There is an approximate particle anti-particle symmetry about this point
and one expects artefacts to be dominant \cite{rind}.  \fig\ref{fig:dense} (right) shows the continuum extrapolated results for
the baryon density as a function of chemical potential, each point being extrapolated from results for 4-7 lattice spacings
\cite{ka4}.  We observe the silver blaze property followed by the onset of nuclear matter, which is steeper for $N_f=2$ than for
$N_f=1$ as expected. Note that the onset transition happens at $\mu_c\losim m_B$, due to the binding energy between the nucleons.
Note also that the onset transition here is a smooth crossover, in contrast to the first-order phase transition for physical QCD
in nature. This is due to the fact that the binding energy decreases with growing quark mass and for the heavy quarks studied here
is smaller than the temperature realized in the plot. It was confirmed in \cite{ka4} that the first-order behaviour indeed results
for sufficiently large hopping parameters/small quark masses, but in that mass range the effective theory is not yet converging
and terms of higher order in $\kappa$ are necessary to quantitatively reproduce QCD.  

\section{Systematics of the effective theory to order $\kappa^8$}

\begin{figure}[t]
  \centerline{
    \includegraphics[width=0.5\textwidth]{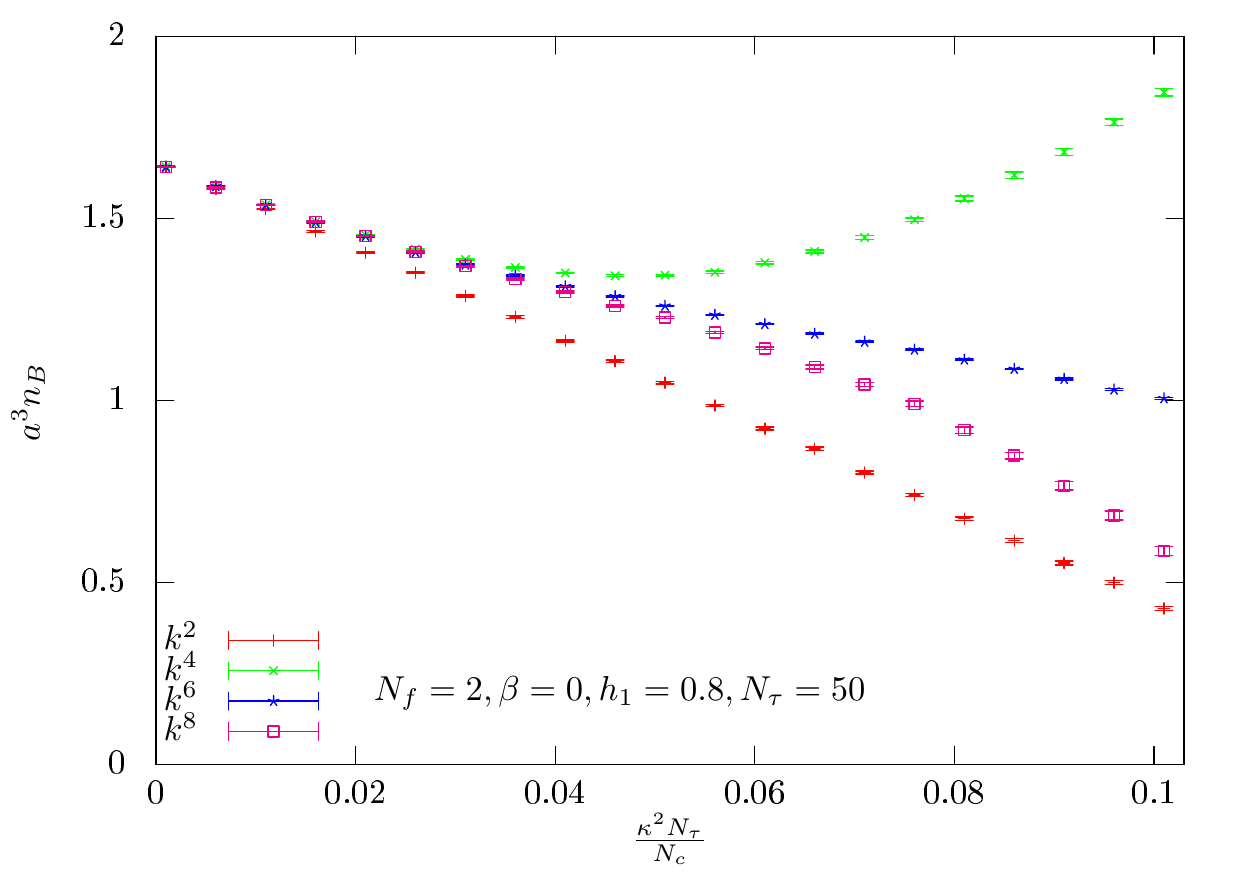}
    \includegraphics[width=0.5\textwidth]{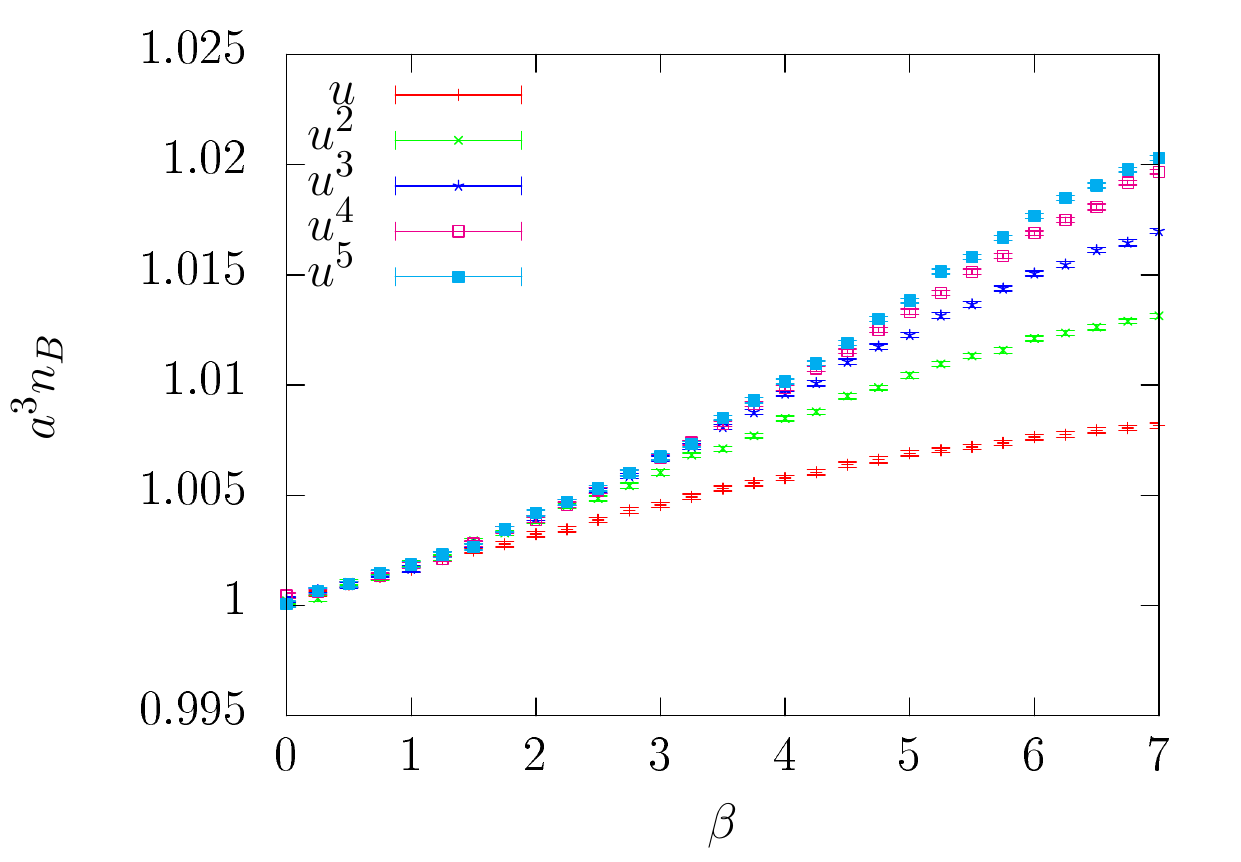}
  }
  \caption[]{Left: Convergence of the baryon density as a function of $h_2 = \frac{\kappa^2 N_t}{N_c}$, computed with 
    effective actions of different orders in the hopping expansion. Right: Convergence in $u$.}
  \label{fig:conv}
\end{figure}

An important question is for which parameter regions the effective theory is valid. Since both the character as well as the
hopping expansion provide convergent series within their radius of convergence, we are able to self-consistently check this by
comparing physical observables at different orders of the effective action. Here we are interested in the cold and dense  region
around the onset of nuclear matter and our observable of choice is the baryon number.  \fig~\ref{fig:conv} (left) tests the
hopping expansion in the strong coupling limit by comparing results obtained with effective actions of increasing order in
$\kappa$.  One observes clearly how adjacent orders stay together for larger values of the coupling $h_2$ as the order is
increased, thus extending the range where our effective action is reliable.  \fig\ref{fig:conv} (right) shows the same exercise
for the largest $h_2\sim 0.1$ considered here, this time increasing the orders of the character expansion. We observe good
convergence up to $\beta\sim 7$, which is a sufficiently weak coupling to allow for continuum extrapolations. It is interesting to
note that the convergence properties are not determined by the size of the expansion parameters alone.  Even though the
$u(\beta)$-values far exceed the $\kappa$-values employed in the figures, convergence in $u(\beta)$ appears to be faster.

The gain in convergence region can be exploited in two ways. Firstly, at fixed temperature and quark masses it allows for the use
of finer lattices, which can be employed in a continuum extrapolation. \fig\ref{fig:dense} (middle) shows results from our
previous simulations obtained with the $\kappa^4$-action as well as new ones with the $\kappa^8$-action at two values of
$\mu>\mu_c$. The baryon density just about reaches the domain with leading cut-off effects linear in $a$, as is expected for
Wilson fermions. The break-off from this behaviour for finer lattices is due to truncation errors and indicates the limit of
validity of the effective action.  The new data generated with the $\kappa^8$-action indeed smoothly extends the linear section
towards the continuum limit. We conclude that our hopping expansion is systematic and controlled, with additional orders in the
action allowing for simulations on finer lattices. For sufficiently heavy masses a continuum extrapolation appears possible.  

\begin{figure}[t]
  \centerline{
    \includegraphics[width=0.45\textwidth]{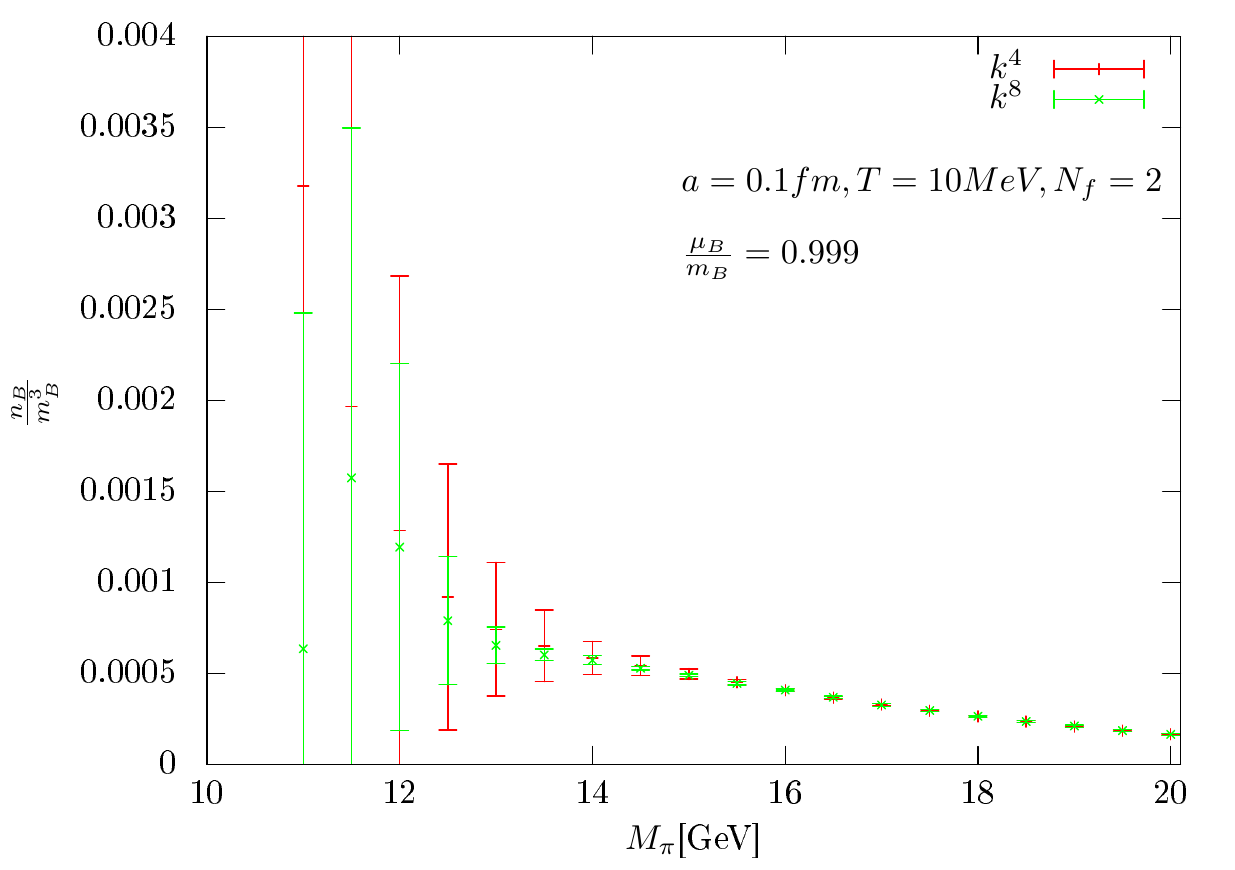}
    \includegraphics[width=0.45\textwidth]{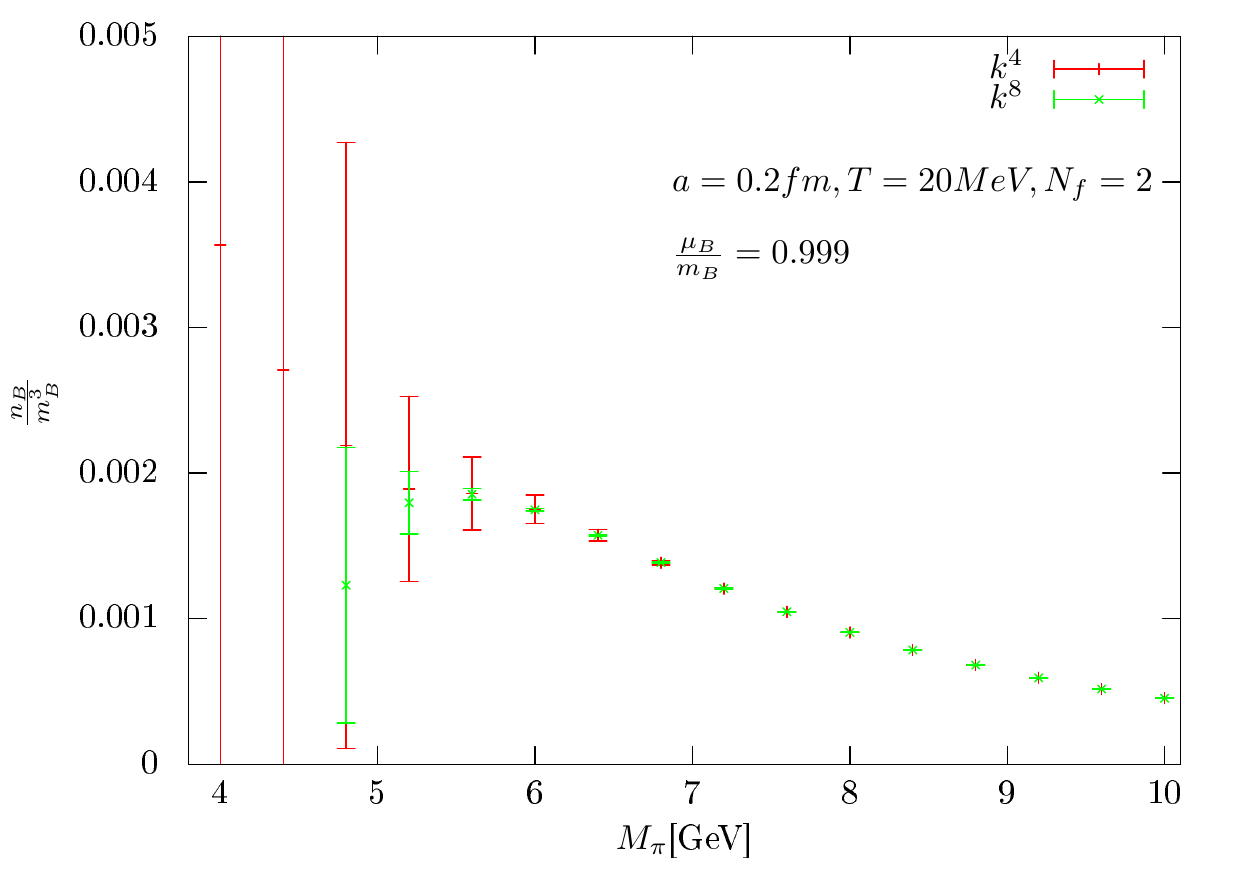}
  }
  \caption[]{Baryon number density as a function of pion mass.}
  \label{fig:mass}
\end{figure}

A second way to benefit from the additional orders in the hopping expansion is to keep the lattice spacing fixed and study smaller
masses. This is shown in \fig\ref{fig:mass} for two different lattice spacings. The error bars in these plots are systematic and
give the difference between results obtained by the action to the highest two orders in the hopping expansion. Growing error bars
thus indicate the loss of good convergence.  As expected, increasing orders allow for smaller quark masses, and so do coarser
lattices.  However, the gain in mass range per additional order in the hopping expansion is too small to envisage an extension to
the physical quark masses of QCD at the present stage.

\section{Analytic evaluation of the effective theory}

So far we have used Monte-Carlo and Complex Langevin simulations to simulate the effective theory on a 3d lattice and numerically
calculate the remaining integral over Polyakov loops. However, restricted to some finite order in the expansion parameters
$\kappa$ and $u(\beta)$, these integrals can also be carried out analytically. In particular, since the effective couplings are
small and correspond to power series in the original couplings, a perturbative evaluation should have good convergence behaviour,
as noted before \cite{bergner,ka4}.  In order to get the correct thermodynamic limit we now employ a linked cluster expansion, and
thus calculate the free energy directly rather than going through the partition function. A review of the standard linked cluster
expansion can be found in \cite{Wortis:1980}.  One should note however that the standard linked cluster expansion tailored to spin
models deals with nearest neighbour interactions. Our action on the other hand contains $n$-point interactions at all distances
and with all possible geometries. It can then be mapped into a spin model, whose partition function thus takes the more
generalised form
\begin{align}
  \mathcal{Z} = \int \prod_{x,i} \mathrm{d} &\phi_i(x) \, \exp \bigg\{ -S_0[\phi_i] %
    - \frac{1}{2!} \sum_{x,y} v_{ij}(x,y) \phi_i(x) \phi_j(x)\nonumber\\
  &-\frac{1}{3!}\sum_{x,y,z} u_{ijk}(x,y,z) \phi_i(x) \phi_j(y) \phi_k(z) + \dots \bigg\}.
\end{align}
In our case the effective theory is fully encoded in the coupling constants $v_{ij}(x,y), u_{ijk}(x,y,z), \dots$, which themselves are
series in the expansion parameters. The free energy can hence be rewritten as
\begin{align}
  \mathcal{W}[v,u,\dots] = \Bigg[&\exp\bigg(\frac{1}{2!}\sum_{x,y}\sum_{i,j}v_{ij}(x,y)
    \frac{\delta}{\delta\tilde{v}_{ij}(x,y)}\bigg) \nonumber\\
  &\exp\bigg(\frac{1}{3!}\sum_{x,y,z}\sum_{i,j,k}u_{i,j,k}(x,y,z)
    \frac{\delta}{\delta\tilde{u}_{i,j,k}(x,y,z)}\bigg) \nonumber\\
  & \hspace{3.4cm}\cdots\hspace{3.4cm} \Bigg] \mathcal{W}[\tilde{v},\tilde{u},\dots]
    \Bigg|_{\substack{\tilde{v} = 0\\\tilde{u} = 0\\\dots}},
\end{align}
and with the expressions for the couplings at hand, the 3d theory can be evaluated analytically in a systematic, well defined way
by carrying out the higher order derivatives and summing over the set of topologically invariant terms.  Although this approach is
valid, it lacks the elegance of the graphical techniques of the linked cluster expansion.  However, because all of the terms
given in \eq\eqref{eq:effective_action} can be embedded on a square lattice, one can use another embedding scheme to simplify the
calculations. This scheme will be introduced next.

\section{Graphical methods}

The linked cluster expansion graphs constitute all contributions to the free energy in the thermodynamic limit up to a certain
order. Therefore by embedding the terms from the effective action onto all connected skeleton graphs from the linked cluster
expansion with two-point interactions, we can calculate the free energy of the effective theory.  This procedure is most easily
demonstrated with an example.

Consider for now an effective theory consisting of two terms, a nearest neighbour term, and a "wedge"
\begin{equation}
  S =
  \;a \sum_{\mathrm{dof}} \; \begin{midtikzpicture}
    \node[dot] (n1) {};
    \node[dot] (n2) at ([shift={(270:.75)}] n1) {}
      edge[thick] (n1);
  \end{midtikzpicture}
  \;+b\; \sum_{\mathrm{dof}}  \: \begin{midtikzpicture}
    \node[dot,blue] (n1) {};
    \node[dot,blue] (n2) at ([shift={(60:{sqrt(2/3)})}] n1) {}
      edge[thick,blue] (n1);
    \node[dot,blue] (n3) at ([shift={(300:{sqrt(2/3)})}] n2) {}
      edge[thick,blue] (n2);
  \end{midtikzpicture} .
\end{equation}
To calculate the free energy, we can embed these terms onto the linked cluster graphs, calcu\-lating symmetry factors and lattice
embeddings for every embedding on every graph. For example embedding onto the fourth order graph
\begin{equation}
  g_8 = \begin{midtikzpicture}
    \node[inner] (n1) {};
    \node[inner] (n2) at ([shift={(60:.75)}] n1) {}
      edge[bend right=30] (n1);
    \node[inner] (n3) at ([shift={(300:.75)}] n2) {}
      edge[bend left=45] (n2)
      edge[bend right=45] (n2);
    \node[inner] (n4) at ([shift={(60:.75)}] n3) {}
      edge[bend left=30] (n3);
  \end{midtikzpicture}, \hspace{1cm} \text{ symmetry: } 4
\end{equation}
we get the following contribution to $\mathcal{W}$
\begin{equation}
  \mathcal{W}_{g_8} =
  \;\frac{a^4 \Lambda}{4} \; \begin{midtikzpicture}
    \node[dot] (n1) {};
    \node[dot] (n2) at ([shift={(60:.75)}] n1) {}
      edge[thick,bend right=30] (n1);
    \node[dot] (n3) at ([shift={(300:.75)}] n2) {}
      edge[thick, bend left=45] (n2)
      edge[thick, bend right=45] (n2);
    \node[dot] (n4) at ([shift={(60:.75)}] n3) {}
      edge[thick, bend left=30] (n3);
  \end{midtikzpicture}
  +a^2 b \Lambda  \; \begin{midtikzpicture}
    \node[dot,blue] (n1) {};
    \node[inner] (n2i) at ([shift={(60:.75)}] n1) {};
    \node[outer,blue] (n2o) at ([shift={(60:.75)}] n1) {}
      edge[thick, blue, bend right=30] (n1);
    \node[inner] (n3i) at ([shift={(300:.75)}] n2) {};
    \node[outer,blue] (n3o) at ([shift={(300:.75)}] n2) {}
      edge[thick, bend left=45] (n2o)
      edge[thick, bend right=45,blue] (n2o);
    \node[dot] (n4) at ([shift={(60:.75)}] n3) {}
      edge[thick, bend left=30] (n3);
  \end{midtikzpicture}
  +\frac{a^2 b \Lambda}{2}  \; \begin{midtikzpicture}
    \node[dot] (n1) {};
    \node[inner] (n2i) at ([shift={(60:.75)}] n1) {};
    \node[outer,blue] (n2o) at ([shift={(60:.75)}] n1) {}
      edge[thick,bend right=30] (n1);
    \node[inner] (n3i) at ([shift={(300:.75)}] n2) {};
    \node[outer,blue] (n3o) at ([shift={(300:.75)}] n2) {}
      edge[thick, bend left=45,blue] (n2o)
      edge[thick, bend right=45,blue] (n2o);
    \node[dot] (n4) at ([shift={(60:.75)}] n3) {}
      edge[thick, bend left=30] (n3);
  \end{midtikzpicture}
  +\frac{b^2 \Lambda}{2}  \; \begin{midtikzpicture}
    \node[dot,blue] (n1) {};
    \node[inner,blue] (n2i) at ([shift={(60:.75)}] n1) {};
    \node[outer,blue] (n2o) at ([shift={(60:.75)}] n1) {}
      edge[thick, blue, bend right=30] (n1);
    \node[inner,blue] (n3i) at ([shift={(300:.75)}] n2) {};
    \node[outer,blue] (n3o) at ([shift={(300:.75)}] n2) {}
      edge[thick, bend left=45,blue] (n2o)
      edge[thick, bend right=45,blue] (n2o);
    \node[dot,blue] (n4) at ([shift={(60:.75)}] n3) {}
      edge[thick, blue, bend left=30] (n3);
  \end{midtikzpicture} .
\end{equation}
The denominators give the overall symmetry factor of the embedding and $\Lambda$ is the lattice embedding of the graph itself. For
$d$ dimensional square lattices $\Lambda=(2d)^3$. After the embeddings we are left with overlapping nodes at individual
lattice points which gives integrals on the form
\begin{equation}
  \int \mathrm{d} W \; \det \big[Q_{\mathrm{stat}}\big]^{2 N_f} W_{n_1,m_1}^{k_1} \cdots
   W_{n_N, m_N}^{k_N}\;,
\end{equation}
which are analytically calculable for $\sum_i n_i k_i \leq 2 N_f$ using for example Polyakov loop integral lookup tables. This means
that for a system with $2$ degenerate quark flavours we can carry out the calcu\-lation up to order $\kappa^8$. The full result for
the free energy is too lengthy to be included here, and will be published elsewhere in the near future.

\section{Comparison with numerics}

\begin{figure}[t]
  \centerline{
    \includegraphics[width=0.5\textwidth]{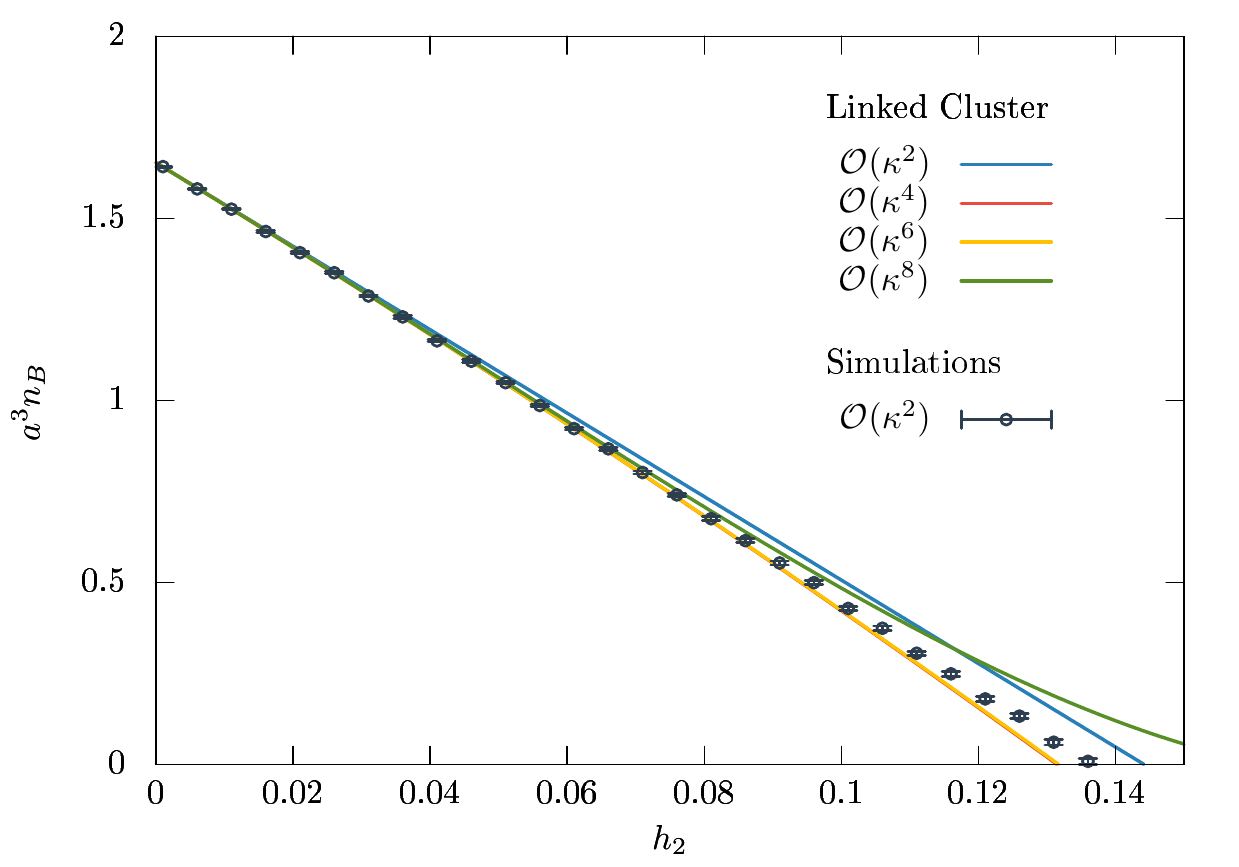}
    \includegraphics[width=0.5\textwidth]{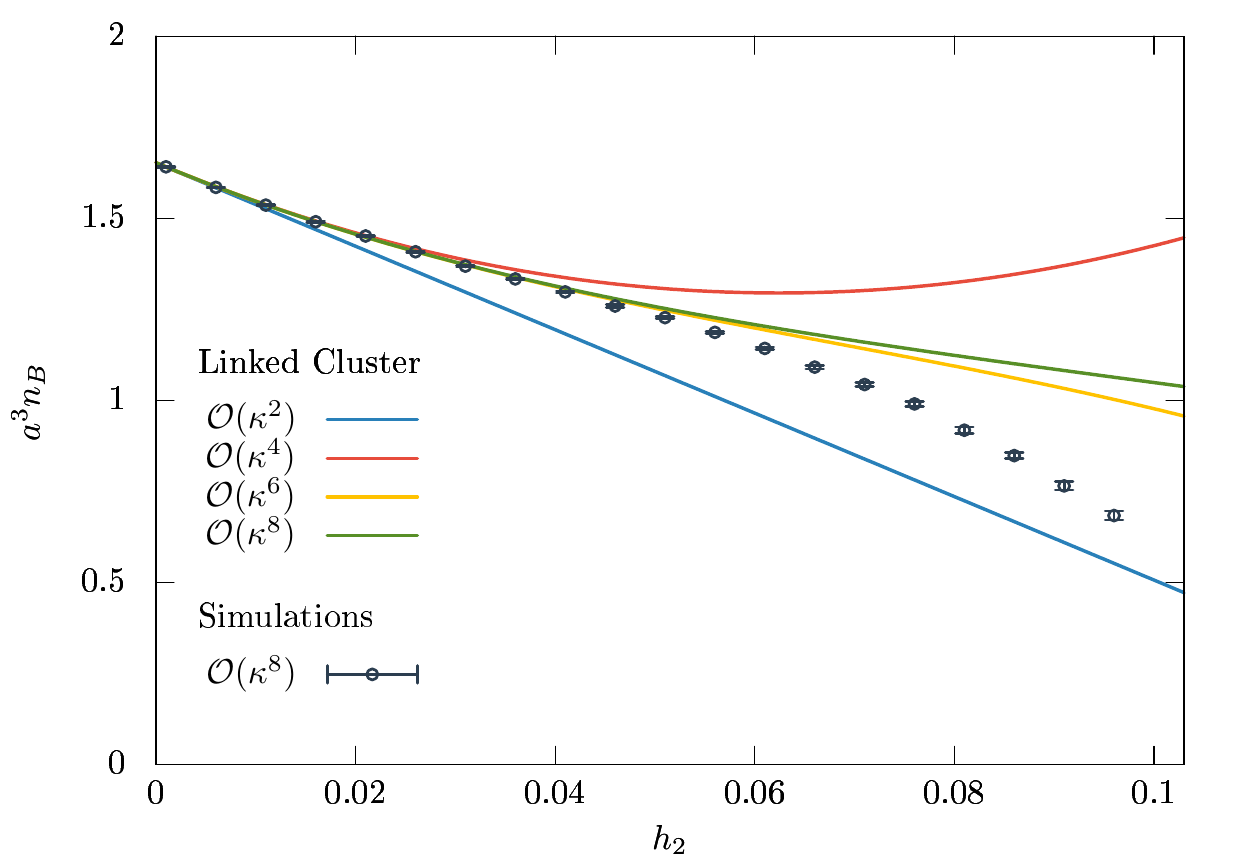}
  }
  \caption[]{
    Linked cluster expansion of the $\kappa^2$ (left) and $\kappa^8$ (right) actions for
    $\beta=0$ and $h_1=0.8$.}
  \label{fig:an_conv}
\end{figure}

With an analytic expression for the free energy at hand, various bulk thermodynamic quantities can be calculated and the results
compared to those obtained with numerics.  It should be stressed that the analytical results emerged from a two stage expansion, first the
effective action is calculated to a fixed order. Second, the expanded action is used to compute e.g. the free energy to a specific
order. This implies that we have for example a $\mathcal{O}\big(\kappa^8\big)$ linked cluster expansion for the
$\mathcal{O}\big(\kappa^2\big)$ action, which we expect to reproduce the $\mathcal{O}\big(\kappa^2\big)$ numerical results.  A
comparison between the various orders of the linked cluster expansion is presented in \fig\ref{fig:an_conv}. Expanding the
$\kappa^2$ action is shown to the left, while the $\kappa^8$ action is used in the right plot.  From the figure one can see that
the $\kappa^2$ action converges much quicker than the $\kappa^8$ action.  This is only natural as the cluster expansion to order
$\kappa^4$ only contains effective action terms to the same order, and the expansion thus have to "catch up" with the order at
hand.

\begin{figure}[t]
  {\centering
    \includegraphics[width=.5\textwidth]{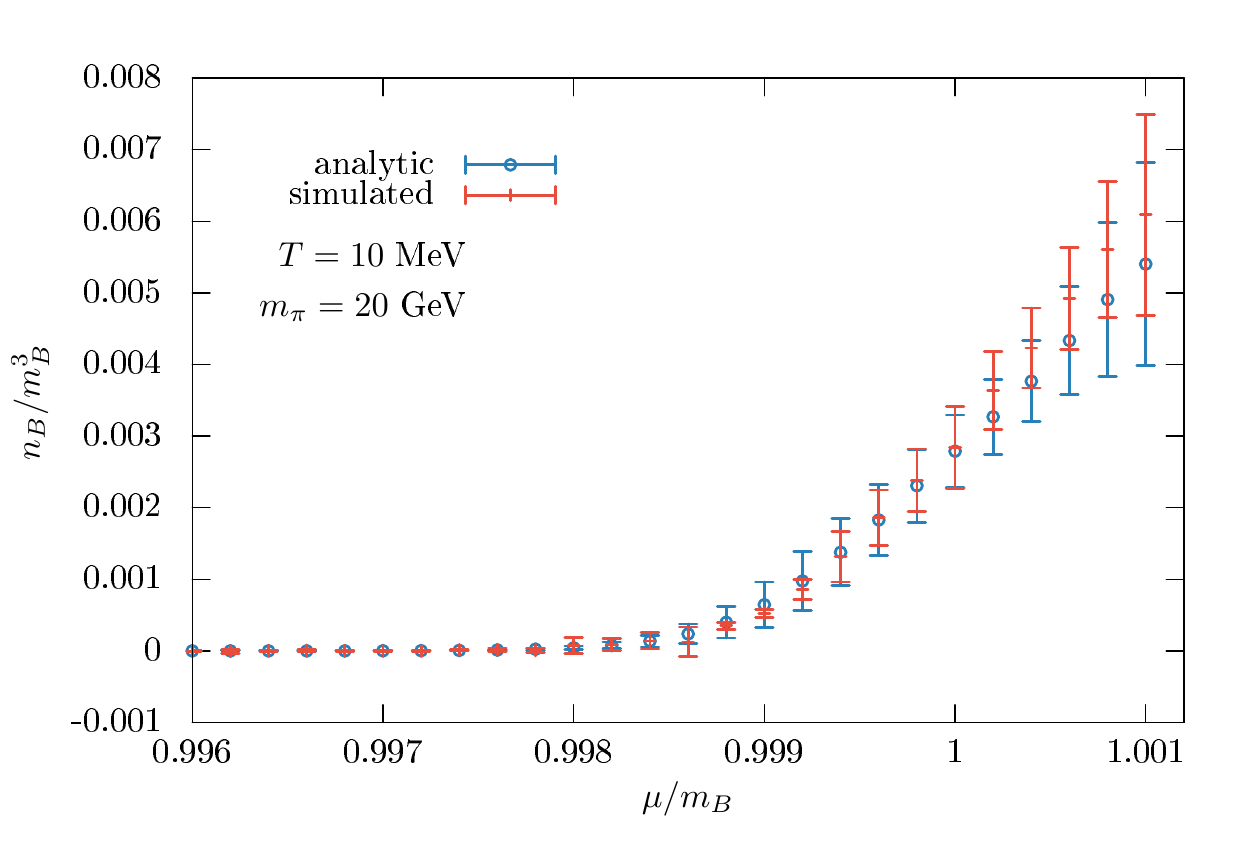}
    \includegraphics[width=.5\textwidth]{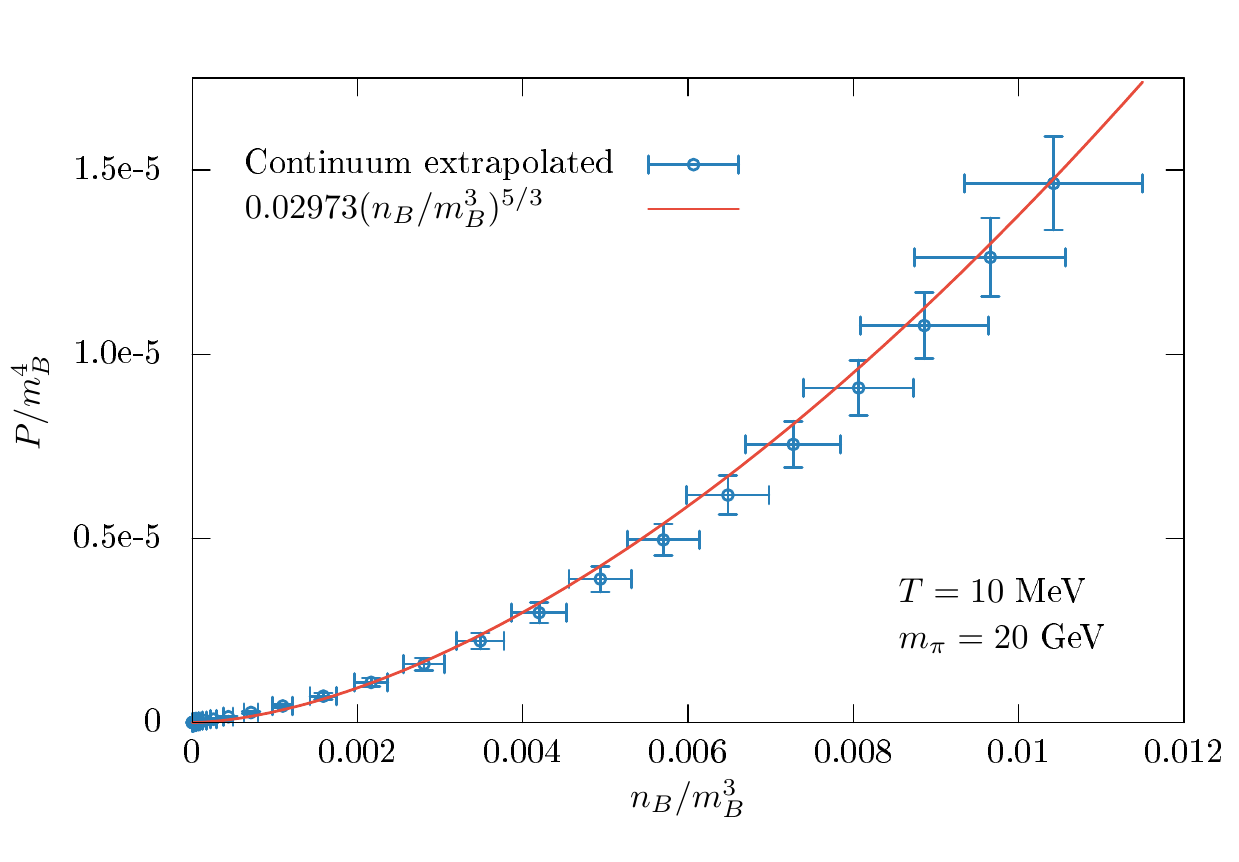}
  }
  \caption{Continuum extrapolated results from the analytic calculation. Left: Baryon density as a function of baryon chemical
    potential, compare to \fig\protect\ref{fig:dense}. Right: The equation of state in the cold and dense regime with a fitted curve.}
  \label{fig:cont_eos}
\end{figure}

A linked cluster expansion is done for a fixed lattice spacing. By applying the same procedure to different lattices it is also
possible to carry out continuum extrapolations. In \fig\ref{fig:cont_eos} (left) it is demonstrated that this gives remarkably
similar results to the numerical ones. In both cases error bars include a systematic error, measured as the difference between the
two highest order effective actions along with an estimate of the continuum extrapolation error. \fig\ref{fig:cont_eos} (right)
shows a continuum extrapolated equation of state for the dense nuclear matter. It is interesting to note that when the curve is
fitted to a power law, the exponent suggests that heavy dense matter behaves as a non-relativistic free gas of fermions.  The
origin of this behaviour will be the subject of further study and we believe that analytical equations will be helpful in this
endeavour as one can toggle the interconnected degrees of freedom individually to obtain insight into the underlying processes at
work.

\section{Chain resummation} \label{sec:chain}

So far we have managed to reproduce most of the simulated results with analytical calculations.  In this section we will present a
resummation scheme for the analytic approach which even extends its reach beyond that of the numerical methods. First let us
demonstrate the pattern with a small example before moving on to the full resummation. Consider for now the following four terms
from \eq\eqref{eq:effective_action},
\begin{equation}
  h_2 N_f \sum_{\mathrm{dof}} \; \begin{midtikzpicture}
    \node[circ] (n1) {1};
    \node[circ] (n2) at ([shift={(270:.75)}] n1) {1}
      edge[thick] (n1);
  \end{midtikzpicture}, \;
  - h_2^2 N_f\sum_{\mathrm{dof}}  \: \begin{midtikzpicture}
    \node[circ] (n1) {1};
    \node[circ] (n2) at ([shift={(60:{sqrt(2/3)})}] n1) {1}
      edge[thick] (n1);
    \node[circ] (n3) at ([shift={(300:{sqrt(2/3)})}] n2) {1}
      edge[thick] (n2);
  \end{midtikzpicture}, \;
  h_2^3 N_f\sum_{\mathrm{dof}} \: \begin{midtikzpicture}
    \node[circ] (n1) {1};
    \node[circ] (n2) at ([shift={(60:{sqrt(2/3)})}] n1) {1}
      edge[thick] (n1);
    \node[circ] (n3) at ([shift={(300:{sqrt(2/3)})}] n2) {1}
      edge[thick] (n2);
    \node[circ] (n4) at ([shift={(60:{sqrt(2/3)})}] n3) {1}
      edge[thick] (n3);
  \end{midtikzpicture}, \;
  - h_2^4 N_f\sum_{\mathrm{dof}} \: \begin{midtikzpicture}
    \node[circ] (n1) {1};
    \node[circ] (n2) at ([shift={(60:{sqrt(2/3)})}] n1) {1}
      edge[thick] (n1);
    \node[circ] (n3) at ([shift={(300:{sqrt(2/3)})}] n2) {1}
      edge[thick] (n2);
    \node[circ] (n4) at ([shift={(60:{sqrt(2/3)})}] n3) {1}
      edge[thick] (n3);
    \node[circ] (n5) at ([shift={(300:{sqrt(2/3)})}] n4) {1}
      edge[thick] (n4);
  \end{midtikzpicture} .
\end{equation}
It is easy to see a common pattern appearing from these terms.  Each term extends the length of the chain by one node while
maintaining a common prefactor. Looking at the equations we see that every link in the chain adds a factor $h_2 W_{2,1}$ to the
term, along with the necessary spatial geometry. One can check with the terms up to order $\kappa^8$ in
\eq\eqref{eq:effective_action} that this holds not only for the simple chain shown above, but for all terms in the action with a
singly connected node, meaning nodes that correspond to a factor of $W_{1,1}$.  The "chain" resummation scheme thus corresponds to
the substitution
\begin{equation}
  W_{1,1}(x) \to W_{1,1}(x) \sum_{n=0}^{\infty} \mathcal{G}(\{x_n\}) \prod_{i=1}^{n} (-h_2) W_{2,1}(x_i),
\end{equation}
where $\mathcal{G}(\{x_n\})$ contains the geometry of the chain. We will not go into more detail about the correctness of the
chain resummation here as it is quite involved. A more satisfactory argument based on the specifics of the effective 3d theory
will be given in a separate publication.

\begin{figure}[t]
  {\centering
    \includegraphics[width=.5\textwidth]{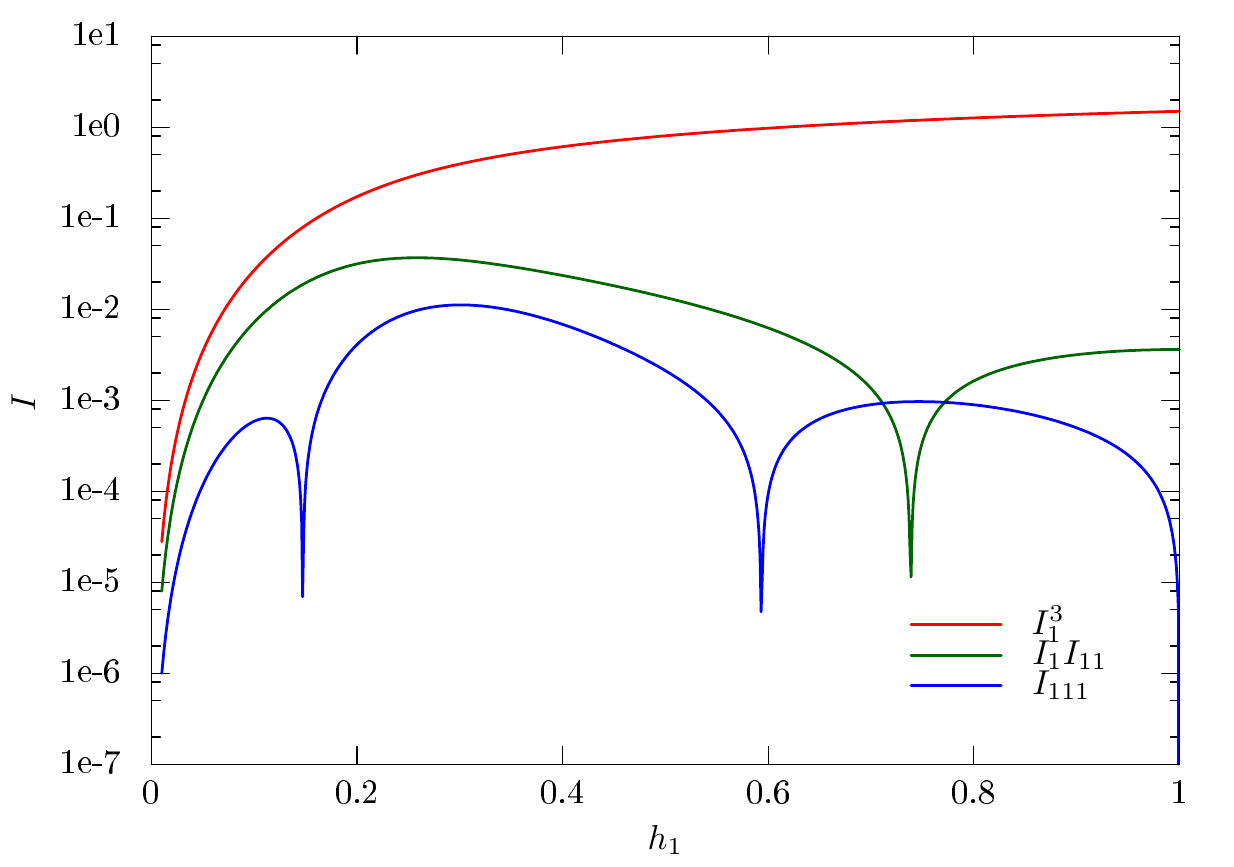}
    \par}
  \caption{Value of the integrated nodes where the red curve comes from three separate nodes, the green curve from 
    the case where two of the three nodes overlap and finally the blue curve from the case where all three nodes overlap.}
  \label{fig:overlapping}
\end{figure}

While we now have a resummation to all orders in $\kappa$ for every order in the effective theory, evaluating the final gauge
integral is still impossible. This is because we need to sum over all geometries for all the terms in
the resummation, which is inaccessible. To proceed, an additional constraint must be introduced, namely that only embeddings with
the same basic geometry as the chain itself is included.  This implies that all nodes of the chain will be at separate
lattice points, and an $n+1$ long chain will result in the following integral
\begin{align}
  \bigg( (2d) h_2 \int \mathrm{d} W \; \det\big[Q_{\mathrm{stat}}\big]^{2 N_f} W_{2,1} \bigg)^n 
  (2d) h_2 \int \mathrm{d} W\; \det&\big[Q_{\mathrm{stat}}\big]^{2 N_f} W_{1,1}  \nonumber\\
  &\equiv (2d h_2)^{n+1} I_{11} I_{21}^n.
\end{align}
and thus the full chain will give
\begin{equation}
  (2d) h_2 I_{11} \sum_{n=0}^{\infty} (-(2d) h_2)^n I_{2,1}^n = \frac{(2d) h_2 I_{11}}{1 + (2d)h_2 I_{21}}.
\end{equation}
Carrying out this resummation also introduces a small error. This arises because the embedding factor of $(2d)$ is too large, and
the linked cluster expansion relies on cancellations from other same order graphs to correct this. These would be graphs that
overlap with itself as it spans the lattice. One could replace the embedding factor with a more natural one from e.g.  studies of
the self-avoiding walk. For a 3d square lattice the factor would be somewhere between 4 and 5. However, in
\fig\ref{fig:overlapping} one can clearly see that the contributions from self-overlapping terms are subleading, especially as we
increase the baryon chemical potential.

\section{Resummed results}

\begin{figure}[t]
  {\centering
      \includegraphics[width=.5\textwidth]{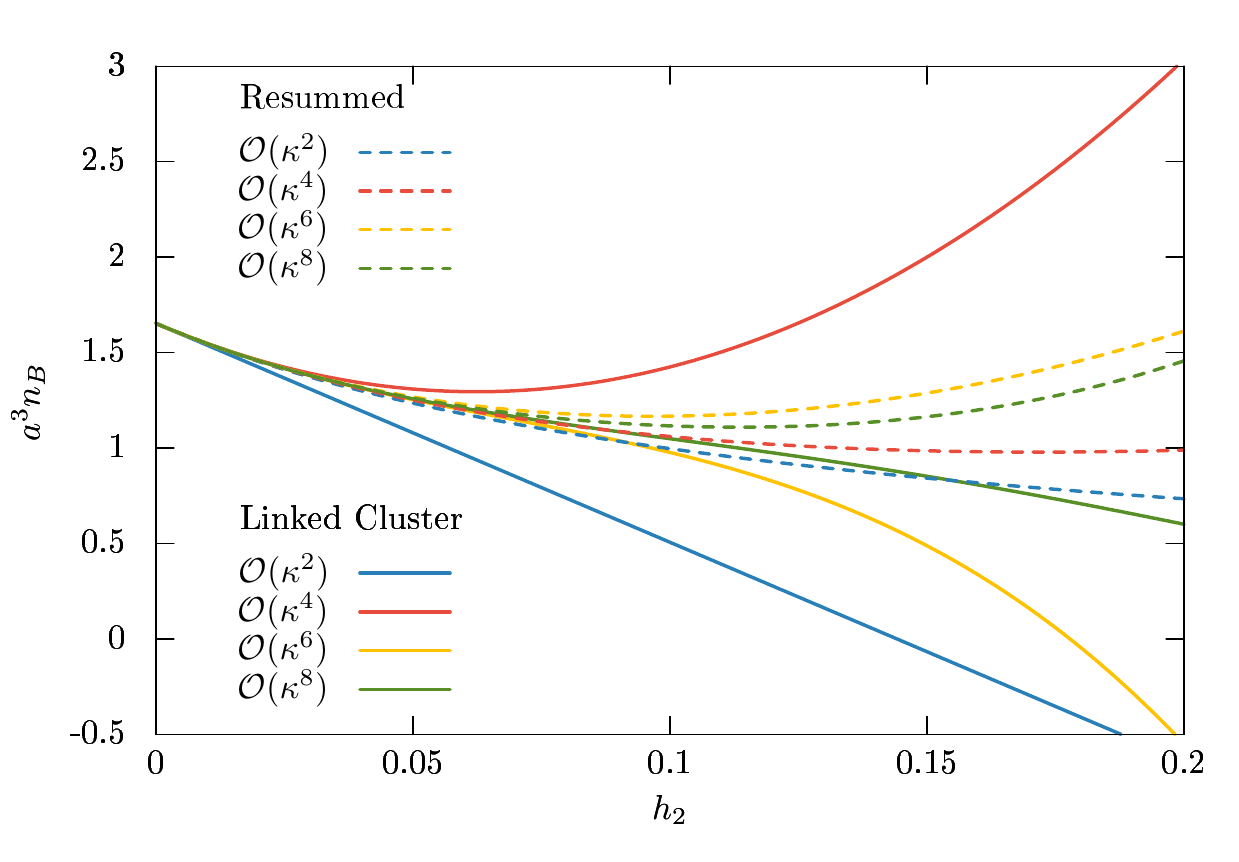}
      \includegraphics[width=.5\textwidth]{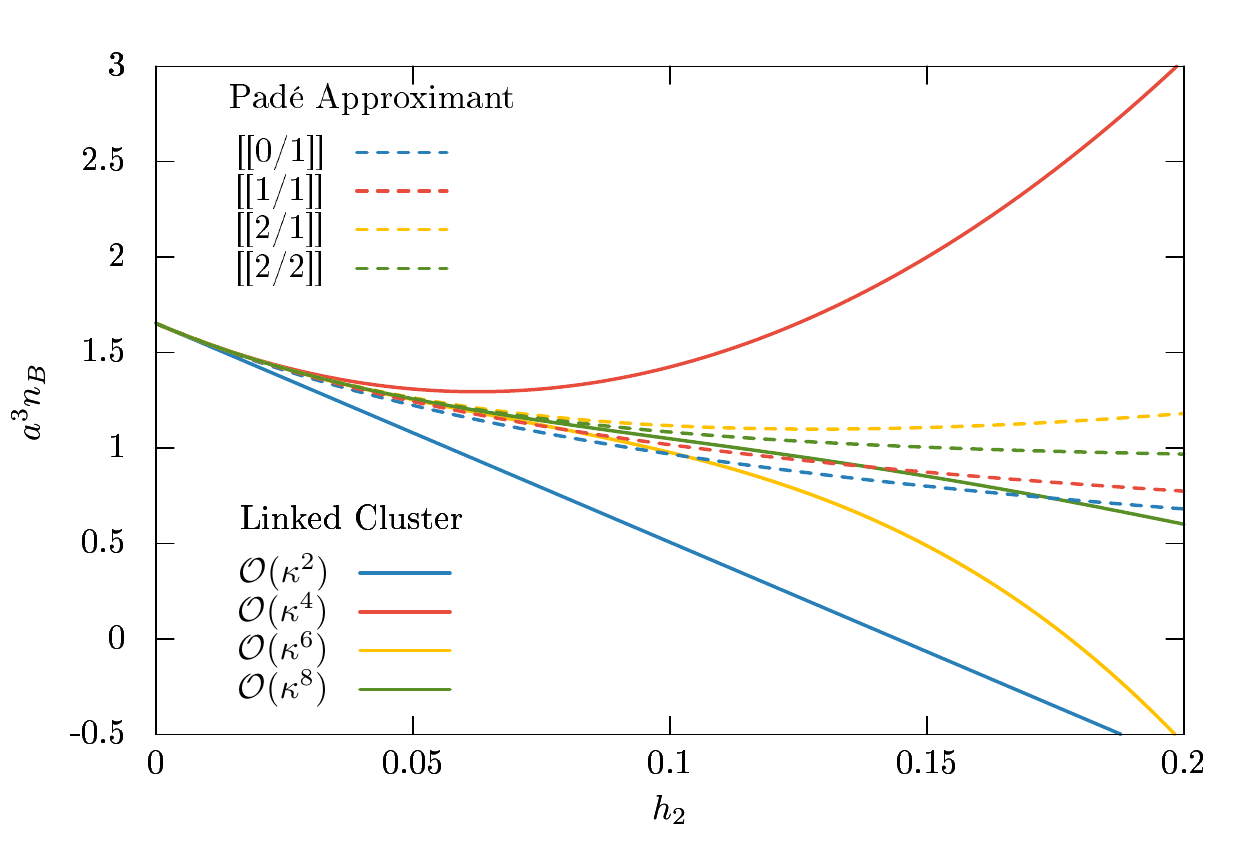}
  }
  \caption{Convergence of the analytically improved 3d theory results. Left: Using the resummation scheme introduced in
    Section~\protect\ref{sec:chain}. Right: Using the near diagonal Pad\'{e} approximants.}
  \label{fig:improved_convergence}
\end{figure}

With the improved analytical results at hand we redo the convergence plot.  In \fig\ref{fig:improved_convergence} (left) the
effects of the resummation are clearly visible, more than doubling the convergence region in $h_2$. On the right of the same
figure we have also provided the result of carrying out a Pad\'{e} approximation in $h_2$, choosing the close to diagonal
approximants. It is not surprising that the Pad\'{e} approximation gives similar improvements, although slightly superior. Both
correspond to resummations giving rational expressions.  The Pad\'{e} is not restricted to a particular class of diagrams and
might therefore predict more of the higher order behaviour. Nonetheless, the similarity between the two results is reassuring as
to the validity of the chain resummation, which originates from the effective theory itself.

\begin{figure}[t]
  {\centering
      \includegraphics[width=.5\textwidth]{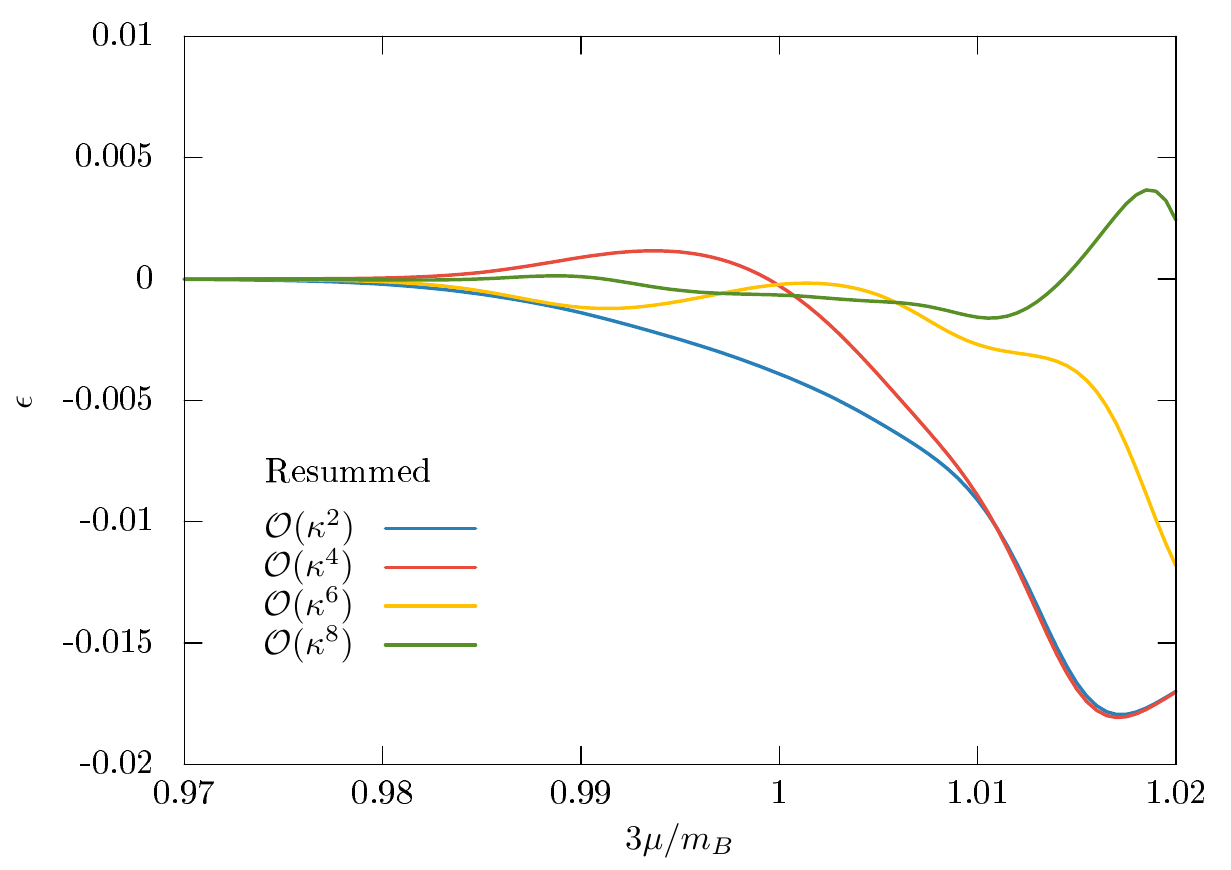}
      \includegraphics[width=.5\textwidth]{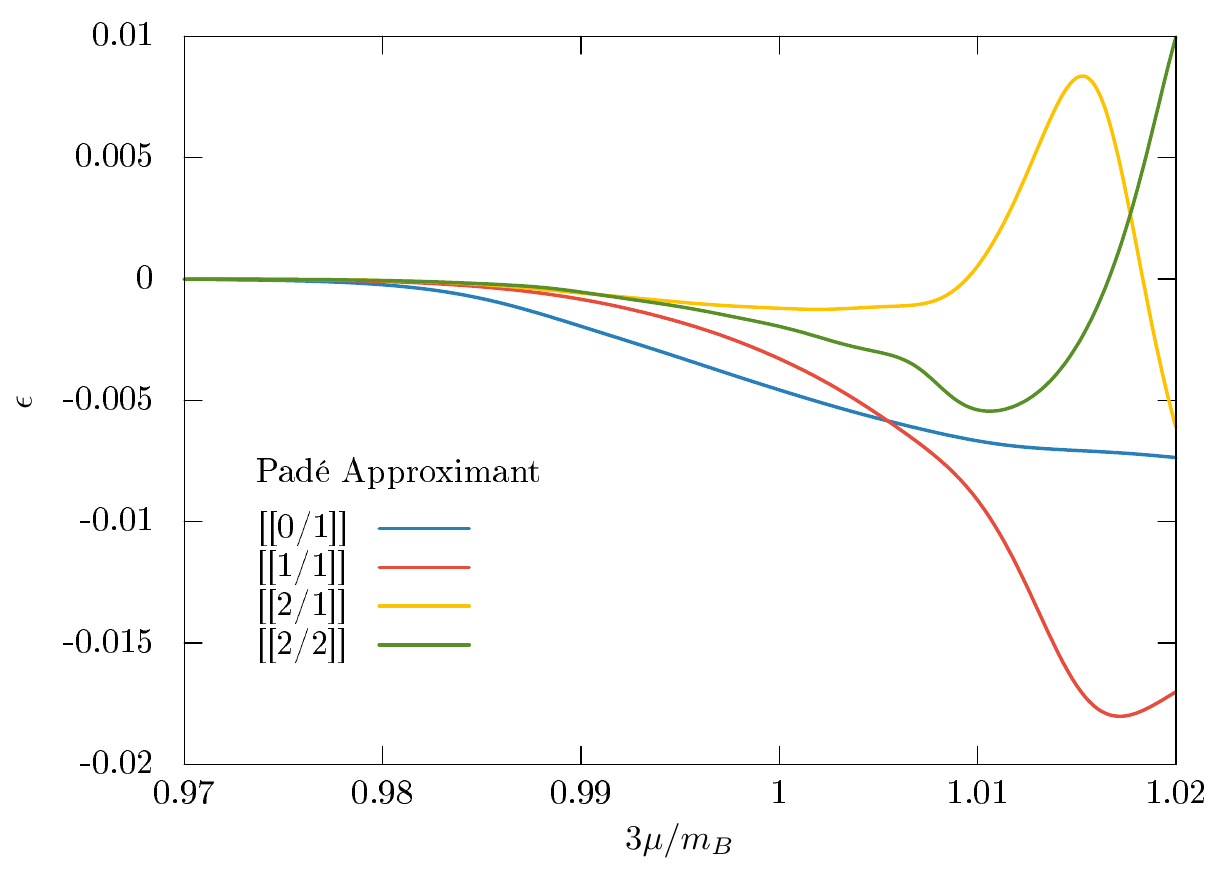}
  }
  \caption{Plots of the binding energy as a function of baryon chemical potential, $\beta=0$, $\kappa=0.08$, $N_t=50$. Left: Using
  the resummation scheme. Right: Using the Pad\'{e} approximants.}
  \label{fig:binding_energy}
\end{figure}

Finally, we use this improved convergence behaviour to study another physics observable.  In \fig\ref{fig:binding_energy} we plot
the binding energy per nucleon, defined by the energy density minus the mass density in the zero temperature limit,
\begin{equation}
  \epsilon \equiv \frac{e - n_B m_B}{n_B m_B}\;.
\end{equation}
This is an important quantity that characterises nuclear matter.  In previous work we have shown numerically and to leading order
in the hopping expansion, that it displays the silver blaze property until the onset transition, where it becomes negative
\cite{ka4}.  We can now extend this study to slightly larger densities. \fig\ref{fig:improved_convergence} shows the binding
energy extracted from both the resummation scheme as well as the Pad\'e approximants to the partition function at various orders.
While quantitative convergence breaks down shortly after the onset transition near $3\mu\sim m_B$, we obtain a new qualitative
result: in higher orders we see the binding energy becoming positive again with growing chemical potential, as is expected from
nuclear physics. A minimum characterising nuclear density appears, which however is not yet settled quantitatively at the
available orders.

\section{Conclusion}

In a continuation of a long term project, we have derived a 3d effective lattice action for the description of thermodynamics of
QCD with heavy quarks through the orders $u^5\kappa^8$ in a combined character and hopping parameter expansion. The effective
action can be simulated by either complex Langevin or, due to the mildness of its sign problem, Metropolis with reweighting.  By
comparing results for observables obtained with effective actions of neighbouring orders, full control over the convergence
properties of the expansion is obtained. The additional orders in the effective action allow for finer lattices and somewhat
lighter quark masses, though an extension to the physically interesting mass region is not in reach at present. By generalising
linked cluster expansion methods from spin models to our effective lattice theory with long distance and many-point couplings, a
fully analytic calculation of thermodynamic functions was achieved that quantitatively agrees with the simulations of the
effective theory.  Finally, we have devised a resummation scheme that sums up infinite chains of graphs analytically, which in
addition includes higher order terms in the effective theory. 

\acknowledgments
This project was supported by the German BMBF, No. 06FY7100 as well as by the Helmholtz International Center for FAIR within the
LOEWE program of the State of Hesse.

\end{document}